\documentclass[fleqn,usenatbib]{mnras}

\usepackage{newtxtext,newtxmath}
\usepackage{blindtext}
\usepackage{caption}
\usepackage{subcaption}
\usepackage{amsmath}
\usepackage{bm}
\usepackage{multicol}
\usepackage{ulem}
\usepackage{xcolor, cancel}

\usepackage[T1]{fontenc}

\DeclareRobustCommand{\VAN}[3]{#2}
\let\VANthebibliography\thebibliography
\def\thebibliography{\DeclareRobustCommand{\VAN}[3]{##3}\VANthebibliography}

\usepackage{graphicx}	% Including figure files
\usepackage{amsmath}	% Advanced maths commands
\usepackage{longtable}
\usepackage{threeparttablex}

\title[Properties of nearly two-hundred new AGNs]{Probing properties of nearly two-hundred new active galactic nuclei}

\author[Ghosh et al.]{
Samrat Ghosh,$^{1,2}$\thanks{E-mail: samratghosh0512@gmail.com}
Samir Mandal,$^{1}$
Sudip Bhattacharyya,$^{3}$
and Shivam Kumaran$^{4}$
\\
$^{1}$Department of Earth and Space Sciences, Indian Institute of Space Science and Technology, Thiruvananthapuram 695547, India\\
$^{2}$Department of Space, Planetary \& Astronomical Sciences \& Engineering, Indian Institute of Technology, Kanpur 208016, India\\
$^{3}$Department of Astronomy and Astrophysics, Tata Institute of Fundamental Research, Mumbai 400005, India\\
$^{4}$Space Applications Center, Ahmedabad 380015, India\\
}

\date{Accepted 2024 September 17. Received 2024 August 22; in original form 2023 September 29}

\pubyear{2024}

\begin{document}
\label{firstpage}
\pagerange{\pageref{firstpage}--\pageref{lastpage}}
\maketitle

\begin{abstract}
We present a comprehensive analysis of the X-ray spectral properties of 198 newly identified active galactic nuclei (AGNs), leveraging archival data from the {\it Chandra} X-ray Observatory. All these AGNs exhibit a powerlaw spectral signature spanning a broad energy range of $0.5-7.0$ keV, characterized by the photon index ($\Gamma$) values ranging from $0.3^{+0.16}_{-0.14}$ to $2.54^{+0.14}_{-0.13}$.
Particularly, 76 of these AGNs display discernible levels of intrinsic absorption, after considering the Galactic absorption.
The column densities associated with this local absorption ($n_{\rm H}^{\rm local}$) are within a range of $\sim 10^{19} - 10^{22}\ {\rm cm^{-2}}$. We study the cosmological evolution of AGNs using the variation of $n_{\rm H}^{\rm local}$ and $\Gamma$ with their estimated redshift. The intrinsic spectral signature did not reveal any significant cosmological evolution; however, a deficit of hard sources at high redshift is possibly intrinsic.
Our sample covers several decades of broadband intrinsic luminosity ($L_{\rm B}^{\rm intr}$) ranging from $4.59^{+0.41}_{-0.41} \times 10^{42}$ to $2.4^{+0.12}_{-0.12} \times 10^{46}\, {\rm erg~s}^{-1}$ with peak at 1.84 redshift.
We also investigate the hardness-luminosity diagram (HLD) to further probe the AGNs. 
We conduct a sanity check by applying our findings to known AGNs, and the results are consistent with our observations. 
\end{abstract}


%%%%%%%%%%%%%%%%%%%%%%%%%%%%%%%%%%%%%%%%%%%%%%%%%%

%%%%%%%%%%%%%%%%% BODY OF PAPER %%%%%%%%%%%%%%%%%%

%%%%%%%%%%%%%%%%%%%%%%%%%%%%%%%%%%%%%%%%%%%%%%%%%%%%%%%%%%%%%%%%%%%%%%%%%%%%%%%%%%%%%%%%%%%%%%%%%%%%%%%%%%%%%%%%%%%%%%%%%%%%
                                                    % Introduction  %
%%%%%%%%%%%%%%%%%%%%%%%%%%%%%%%%%%%%%%%%%%%%%%%%%%%%%%%%%%%%%%%%%%%%%%%%%%%%%%%%%%%%%%%%%%%%%%%%%%%%%%%%%%%%%%%%%%%%%%%%%%%%

\section{Introduction}
Active Galactic Nuclei (AGNs) are the luminous and energetic nuclei of galaxies that produce most of their energy output through actively accreting materials into the central supermassive black hole (SMBH) \citep{Lynden_bell_1969}. 
These exotic objects emit radiation in all wavelengths ranging from radio to $\gamma$-rays \citep{Wilkes_1999}.
Spectral and timing studies across various wavelength bands have played a pivotal role in constraining the distance scales of different emission regions within AGNs. These investigations have contributed to the development of the unification model \cite[][and reference therein]{Antonucci_1993, Urry_and_Padovani_1995, Bianchi_et_al_2012}, which seeks to provide a coherent framework for explaining the diverse observed characteristics of various types of AGNs.

AGNs are classified in various ways, considering their radio, optical, X-ray, and other observations \cite[see][for a complete review]{Padovani_et_al_2017}. 
For example, Type I and Type II AGNs are classified based on the presence of broad and narrow emission lines in the optical spectra \citep[e.g.][]{Stern_&_Laor_2012, Oh_et_al_2015} respectively. In the optical classification of AGNs, Seyfert galaxies are commonly categorized into two main types: Seyfert 1 (Sy1) and Seyfert 2 (Sy2). The presence of broad emission lines in the spectra of Type I AGNs (or Sy1) indicates gas moving at high velocities in the vicinity of an SMBH at the centre of a galaxy. In the X-ray domain, AGNs are often classified \citep[e.g.][]{Garcet_et_al_2007} based on the obscuring column density of material along the line of sight. Type I AGNs are powerful emitters of X-rays and are known for their unobstructed view of the central region, allowing for detailed investigations of their properties and demographics.
 While it is generally expected that unobscured X-ray AGNs correspond to optical Type I AGNs and vice versa, there have been observed discrepancies \citep[e.g.][]{Garcet_et_al_2007, Ordovas_et_al_2017} in some cases.

One of the primary components of AGN is an optically thick and geometrically thin accretion disc, as first postulated by \cite{shakura_sunyaev_1973}. This accretion disc predominantly emits multicolour blackbody radiation in the optical or ultraviolet spectral range. 
However, the X-ray emission is assumed to result from inverse-Compton scattering \citep{Sunyaev_&_Titarchuk1980,Sunyaev_&_Titarchuk1985} of the accretion disc photons by the hot electron cloud or corona \citep[][and references therein]{Liang_1979,Haardt_Maraschi1991,Dovciak_Done_2016} located in the vicinity of the SMBH. 
This dominant and commonly observed Comptonization spectrum from AGNs can often be mimicked by a powerlaw continuum \citep{rybicki_lightman}. 
AGNs X-ray powerlaw photon indices ($\Gamma$) generally lie between $1.5$ and $2.5$ \citep[][and references therein]{Ishibashi_&_Courvoisier_2010}. However, $\Gamma$ varies across different classes of AGNs. Several studies have been done to understand the variation of $\Gamma$ with other AGN properties, e.g., radio loudness, intrinsic obscuration, black hole mass, etc. Studies have shown that radio-loud quasars have generally harder spectra compared to those of radio-quiet quasars \citep[e.g.][etc.]{Lawson_and_Turner_1997, Page_et_al_2005, Piconcelli_et_al_2005, Shaban_et_al_2022}. 
Moreover, \cite{Page_et_al_2004} showed the spectral shape for radio-quiet quasars does not vary with luminosity and black hole mass. \cite{Zdziarski_et_al_1995, Zdziarski_et_al_2000} showed that Sy1 galaxies generally have softer spectra (high $\Gamma$) than those of Sy2 galaxies.

There are other components of the X-ray spectrum, such as a soft excess \citep[e.g.,][]{Porquet_et_al_2004, Piconcelli_et_al_2005, Gierlinski_&_Done_2004}, a reflection hump \citep[][and reference therein]{Fabian_2006, Ricci_et_al_2011, Fausnaugh_et_al_2022}, and a fluorescent Fe$~\rm K\alpha$ line \citep[e.g.,][]{Ricci_et_al_2014,Matt_Fabian_Reynolds_1997,Fabian_et_al_2000}.

This study mainly focuses on the X-ray emission from AGNs. X-rays can penetrate large distances, probing the distant past in the Universe and providing a near complete survey of AGN in this band \citep{Brandt_&_Alexander_2015}. 
Moreover, X-rays from AGNs are not significantly contaminated by the host galaxy emission and provide a clearer view of the nuclear region of active galaxies than that by optical/UV emission \citep[e.g.][]{Moran_et_al_2002, Brandt_&_Alexander_2015}. 
Hence, X-ray studies of AGNs are important to understand the physical nature of such individual sources and their collective behaviour.

This paper studies the X-ray properties (in the $0.5-7.0$ keV band) of 198 AGNs recently identified by \cite{kumaran2023}. Our sample represents a population of X-ray Type I AGNs (hereafter refered to as type-I sources in the context of our sample). Section \ref{sect:sample_select} discusses the sample selection followed by data reduction and spectral modelling in section \ref{sect:data_red_spec_model}. We present our result in Section \ref{sect:result} and discuss the significance of our findings in  \ref{sect:discussion}. Finally, we summarize in Section \ref{sect:summary_and_conclusion}.

%%%%%%%%%%%%%%%%%%%%%%%%%%%%%%%%%%%%%%%%%%%%%%%%%%%%%%%%%%%%%%%%%%%%%%%%%%%%%%%%%%%%%%%%%%%%%%%%%%%%%%%%%%%%%%%%%%%%%%%%%%%%
                                                    % Sample Selection  %
%%%%%%%%%%%%%%%%%%%%%%%%%%%%%%%%%%%%%%%%%%%%%%%%%%%%%%%%%%%%%%%%%%%%%%%%%%%%%%%%%%%%%%%%%%%%%%%%%%%%%%%%%%%%%%%%%%%%%%%%%%%%

\section{Sample Source Selection}
\label{sect:sample_select}
The sample set is selected from the recently identified AGNs in the X-ray energies, employing machine learning techniques outlined in the study by \cite{kumaran2023}, utilizing data from \textit{Chandra} Source Catalog version 2.0 \citep{csc2020}. Their work introduced a class membership probability (CMP) to gauge the credibility of classifying a specific source. CMP assesses the degree of confidence regarding a source's membership in a particular class. A total of 31,824 AGNs were identified with a confidence probability exceeding $99.73\%$. In this study, we have focused on a specific subset of these AGNs and conducted an in-depth analysis of their spectral characteristics. Given that these AGNs have been newly identified in X-ray observations, we have additionally incorporated well-known AGN samples to validate and provide context for our findings.

%%%%%% Sky distribution of the source %%%%%%
\begin{figure}
    \centering
    \includegraphics[width=\linewidth]{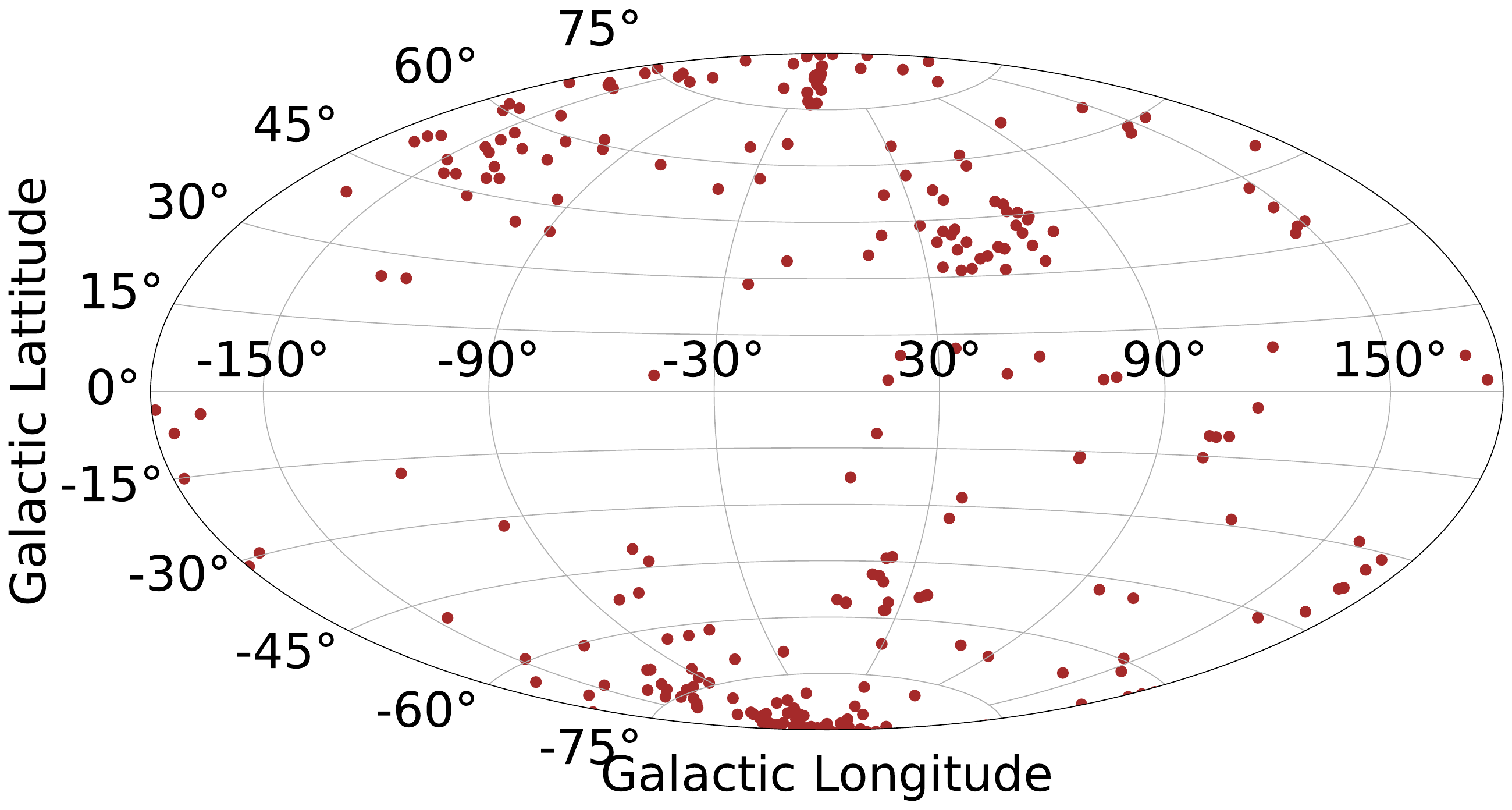}
    \caption{Distribution of new AGNs in the galactic coordinates (see Section \ref{subsect:new_sample_selection}).}
    \label{fig:sky_distr}
\end{figure}

%%%%%%%%%%%%%%%%% Sample selection of new AGNs %%%%%%%%%%%%%%%%%%
\subsection{Sample selection of new AGNs}
\label{subsect:new_sample_selection}
Out of the initial 31,824 AGNs, we have narrowed our focus to a subset of sources, allowing for a more thorough and feasible analysis of their data. 
In making this selection, we have relied on the parameter \texttt{o.src\_cnts\_aper90\_b} from CSC 2.0. This parameter provides the total counts within the source region aperture, encompassing $90\%$ of the total energy of the point spread function (PSF).
We have decided on a minimum threshold of 400 counts within the source region aperture. A detailed analysis  (Appendix~\ref{appendixA}) shows no bias on the source selection. This choice ensures an acceptable signal-to-noise ratio (SNR) in the spectra and a substantial number of sources for our analysis. 
Given the aforementioned cutoff, we have identified 233 AGNs having 264 observations. However, some AGNs exhibit significant off-axis positions in the image, resulting in distorted or blurred PSFs, making it challenging to detect the sources against the background. These particular AGNs were excluded from the analysis following manual examination.
Finally, spectral modelling has been carried out for a subset of 198 AGNs. We have aggregated the spectra from multiple observations conducted within a year to enhance the SNR. Alternatively, we have retained only the observation having the best statistics, typically characterized by the highest number of counts. As a result, each AGN is associated with a single estimate for each spectral parameter under consideration. In Figure \ref{fig:sky_distr}, the source positions are depicted in galactic coordinates to check if our sample selection exhibits any directional bias. Fewer sources are present toward the galactic plane than those in polar directions. This discrepancy arises from the substantial obscuration along the galactic disc. However, it is important to note that sources in this direction are not entirely absent; in fact, a considerable number of sources are found at low galactic latitudes. This observation suggests that our sample is not significantly biased directionally.

\begin{figure}
    \centering
    \includegraphics[width=0.9\columnwidth]{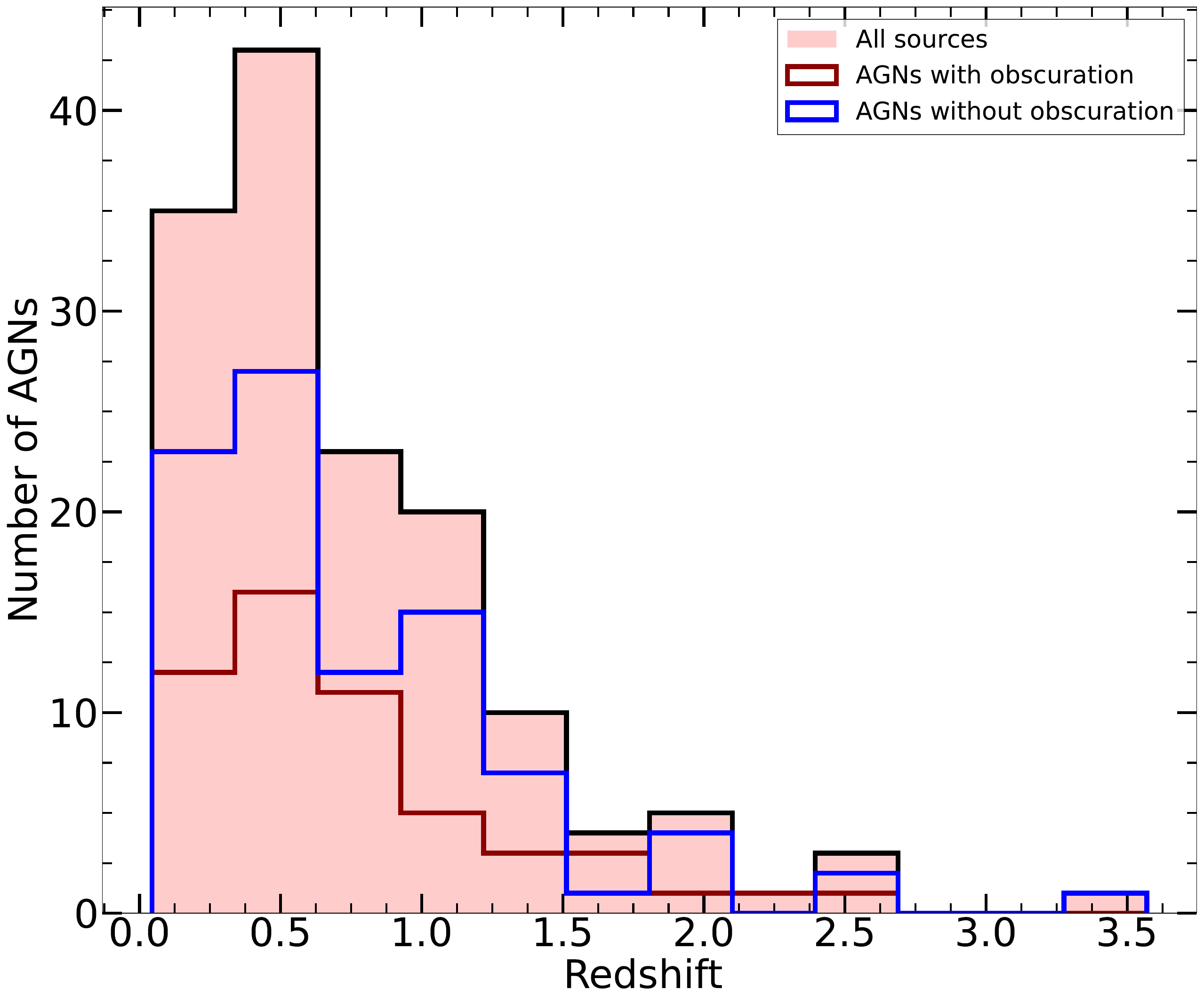}
    \caption{Redshift distribution of new AGNs (see Section \ref{subsect:new_sample_selection}).The red and blue lines indicate the distribution of AGNs with and without local obscuration respectively (see Section \ref{subsect:modeling_new_AGNs} for details.)}
    \label{fig:redshift_distr}
\end{figure}

To acquire redshift measurements, we have utilized the NASA Extragalactic Database (NED) and conducted a cone search with a 5 arcsecond search radius. When multiple sources are present within the search cone for a specific AGN, we select the source with the smallest projected separation. Consequently, we obtain redshift information for 145 AGNs. The distribution of these redshifts is illustrated in Figure \ref{fig:redshift_distr}. Our sample encompasses a broad range of redshift ($z$) values from 0.044 to 3.57. However, the majority of the sources in our study have redshifts below 0.7, while only a limited number of sources have redshifts greater than 2. We also indicate the redshift distribution of AGNs with local obscuration (red line) and without local obscuration (blue line) which are discussed in Section \ref{subsect:modeling_new_AGNs}.

%%%%%%%%%%%%%%%%% Selection of the known sources %%%%%%%%%%%%%%%%%%
\subsection{Sample of known AGNs for a comparison}
\label{subsect:validation_set}
In addition to the sample detailed in Section \ref{subsect:new_sample_selection}, we employ a well-established sample of AGNs in the X-ray to compare our findings. The study by \cite{kumaran2023} utilized the Veron Catalog of AGNs and Quasars 13th Edition (VERONCAT) \citep{veroncat_13th_ed_2010} as their primary resource. We have incorporated their AGN training dataset but primarily focused on a subset comprising known Sy1 galaxies. This emphasis aligns with our sample of newly identified AGNs, which predominantly consists of type-I AGNs, as elaborated in Section \ref{subsect:modeling_new_AGNs}.

The catalogue enlists 15,627 Sy1 galaxies, and upon performing a cross-match with CSC 2.0, we identified 588 matches. A threshold of \texttt{significance=10} from CSC 2.0 is employed to filter this dataset, resulting in 68 AGNs. It is worth noting that all of these AGNs exhibit redshifts below 2. However, in the larger pool of 588 Sy1 galaxies, there exists a subset of 118 AGNs with redshifts exceeding 2. To specifically isolate a subset from this category, we relax the significance threshold to 5. Under this revised criterion, we have identified and selected 53 AGNs with redshifts surpassing 2. By merging these two subsets, one containing AGNs with redshifts less than 2 (68 AGNs) and the other with AGNs exceeding a redshift of 2 (53 AGNs), we have compiled a comprehensive sample set comprising 121 well-known Sy1 galaxies (most of these sources, we believe, are the best representatives of type-I AGNs within our sample of known AGNs). We exclusively utilize \textit{Chandra} observations with the highest exposure time available for each AGN. However, 30 (3 AGNs with $z<2$ and 27 AGNs with $z>2$) sources are excluded due to distorted and/or blurred PSFs. The reason behind more number of smeared sources for $z>2$ is the lower significance level. Consequently, we have retained a total of 91 AGNs in this sample set.

%%%%%%%%%%%%%%%%%%%%%%%%%%%%%%%%%%%%%%%%%%%%%%%%%%%%%%%%%%%%%%%%%%%%%%%%%%%%%%%%%%%%%%%%%%%%%%%%%%%%%%%%%%%%%%%%%%%%%%%%%%%%
                                                    % Data Reduction and Spectral Modeling  %
%%%%%%%%%%%%%%%%%%%%%%%%%%%%%%%%%%%%%%%%%%%%%%%%%%%%%%%%%%%%%%%%%%%%%%%%%%%%%%%%%%%%%%%%%%%%%%%%%%%%%%%%%%%%%%%%%%%%%%%%%%%%
%%%%%% Redshift Distribution %%%%%%
\section{Data Reduction and Spectral modelling}
\label{sect:data_red_spec_model} 
Every AGN within our sample has been observed using the ACIS detector onboard the \textit{Chandra} X-ray Observatory (CXO) (\cite{CXO}). This CCD detector is sensitive in the 0.2 to 10 keV energy range. Nevertheless, the detector's effective area drops drastically below 0.5 keV and above 7.0 keV. Consequently, our analysis considers X-ray data in the energy range of 0.5 - 7.0 keV. 

Our data reduction process adheres to the standard procedure using the \textit{Chandra} Interactive Analysis of Observations (CIAO) package version 4.15 (\cite{CIAO}). The Level 1 event files are obtained from the \textit{Chandra} Data Archive (CDA) \citep{cda}. Subsequently, we have generated Level 2 event files using \texttt{chandra\_repro} script within the CIAO package. The latest calibration database (CalDB) version 4.10.2 is employed in the data reduction process.

%%%%%%%%%%%%%%%%% Spectrum extraction and grouping %%%%%%%%%%%%%%%%%%

For each AGN, we have defined the source region as a circular region of radius (r$_{90}$), which encloses $90\%$ of the energy of the PSF and is calculated using the \texttt{psfsize\_srcs} task within the CIAO package. 
The background region is selected as an annulus of inner radius 1.2 r$_{90}$ and outer radius 2 r$_{90}$. To ensure the appropriateness of these regions, we superimpose them onto the image and verify that there are no sources within the background region. Subsequently, spectra are extracted using the \texttt{specextract} script, which integrates several tasks, namely \texttt{dmextract}, \texttt{mkarf}, and \texttt{mkrmf}. The \texttt{dmextract} task generates appropriate binning along the desired axis (the energy axis for the spectrum). The tasks \texttt{mkarf} and \texttt{mkrmf} are employed to create the Auxiliary Response File (ARF) and the Redistribution Matrix File (RMF), respectively. 
Furthermore, the script \texttt{specextract} also utilizes \texttt{arfcorr} task if the parameter \texttt{correctpsf} is set as `yes'. This task is crucial to apply any PSF correction if required. 
The spectra are grouped with a minimum of 15 or 20 counts per bin (required for $\chi^2$ minimization while model fitting) depending on the situation. The grouping is implemented within the \texttt{specextract} script for AGNs having a single observation. Alternately, for AGNs having multiple epochs of observations, we have combined the spectra, and the resultant spectrum is grouped using \texttt{grppha} task of {\it FTOOLS} package \citep{ftools} in HEASoft version 6.31.1.

%%%%%%%%%%%%%%%%% Spectrum modeling and analysis of new AGNs %%%%%%%%%%%%%%%%%%
%%%%%% local nh distribution %%%%%%
\begin{figure}
    \centering
    \includegraphics[width=0.9\columnwidth]{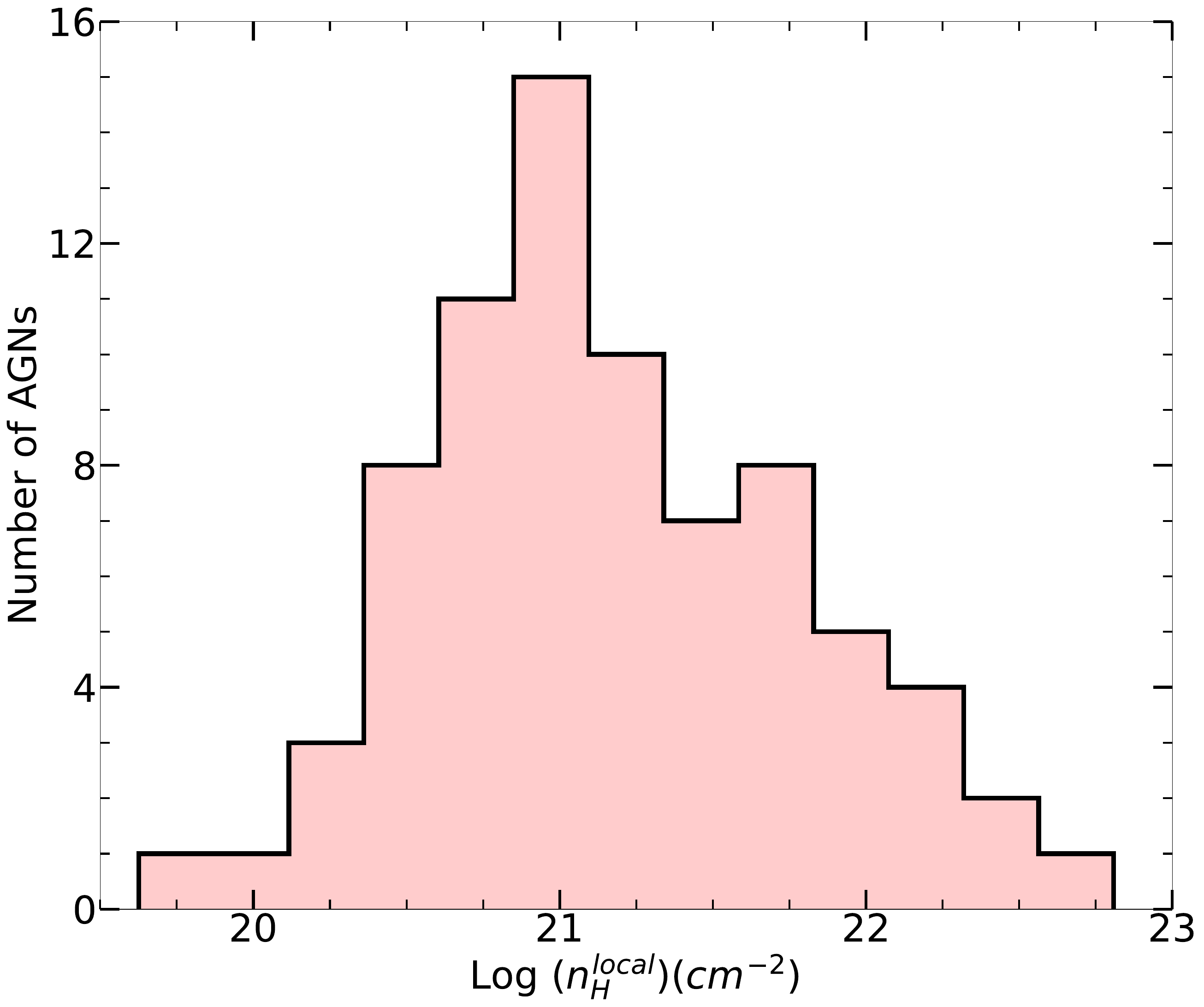}
    \caption{Distribution of local absorption column density ($n_{\rm H}^{\rm local}$) of new AGNs. The absence of bimodality indicates a single population (see Section \ref{subsect:modeling_new_AGNs}).}
    \label{fig:local_nh_distr}
\end{figure}

\subsection{Spectral modelling of new AGNs}
\label{subsect:modeling_new_AGNs}
We have used the {\it XSPEC} package version 12.13.0c \citep{xspec} for spectral modelling. It is widely accepted that the X-ray spectral component in AGN arises from a hot electron cloud or corona due to the inverse Comptonization \citep{Sunyaev_&_Titarchuk1980,Sunyaev_&_Titarchuk1985,x-ray_pow_from_corona_1} of seed optical/UV photons generated from the accretion disc. This process produces a powerlaw continuum emission \citep{rybicki_lightman}. In our sample of new AGNs, a simple absorbed powerlaw model is sufficient for modelling their X-ray spectra. However, in the broader context of AGN X-ray spectroscopy, more complex model components are typically needed to describe the observed data accurately.
Initially, we have modelled the data using only a single absorption component \texttt{tbabs} \citep{tbabs} with a powerlaw i.e., \texttt{tbabs $\times$ powerlaw} keeping the column density parameter ($n_{\rm H}$) free. The line of sight column density ($n_{\rm H}^{\rm MW}$) attributed to absorption due to the Milky Way (MW) can be found from HI $4\pi$ survey \citep{nh} towards the direction of each AGN. In certain instances, there may be additional intrinsic absorption specific to the AGN itself. If the $68\%$ lower confidence level of the $n_{\rm H}$ parameter falls below the value of $n_{\rm H}^{\rm MW}$, we conclude that the AGN is not subjected to any significant local obscuration. 

In our analysis, we have identified evidence of local obscuration in 76 AGNs, where we employ a powerlaw model with two absorption components: \texttt{tbabs $\times$ ztbabs $\times$ powerlaw} to fit the data. The component \texttt{tbabs} represents the line of sight absorption due to the Milky Way, and \texttt{ztbabs} corresponds to the absorption local to the AGN. The column density parameter ($n_{\rm H}^{\rm MW}$) of \texttt{tbabs} is frozen to the aforementioned MW line of sight value, and the local column density $n_{\rm H}^{\rm local}$ (parameter of \texttt{ztbabs}) is allowed to vary freely during our modelling process. 
The redshift parameter of the \texttt{ztabs} model is assigned a value obtained from multi-wavelength observations, as detailed in Section \ref{subsect:new_sample_selection}.
For 28 AGNs in which we lack redshift information and have identified some degree of obscuration, we have adopted a pragmatic approach by setting their redshift, $z=0$. This choice is based on the rationale that, in the absence of specific redshift data, an arbitrary AGN with a lower redshift is more plausible (see Figure~\ref{fig:redshift_distr}) than assigning it a high redshift. 
In principle, we could fix the redshift of these sources to the median value of the distribution shown in Figure \ref{fig:redshift_distr}, However, it may not signify anything useful because the actual redshift can differ on either side of the median value. In contrast, we choose z = 0 for the sources without redshift information because the estimated local column density provides at least a lower bound of the same and the actual value is surely larger than this.

Our initial attempt to estimate the parameter uncertainty using the standard error estimation method within \textit{XSPEC} proved to be inadequate for constraining the lower limit of $n_{\rm H}^{\rm local}$ in certain cases. Consequently, we turned to alternative methods, such as Markov Chain Monte Carlo (MCMC) analysis in \textit{XSPEC}, to obtain a reliable estimate for this parameter in those specific instances.
In this MCMC analysis, the best-fit parameters, determined through the $\chi^2$ statistic, serve as the basis for generating a Gaussian distribution to initiate the walkers. Furthermore, we have defined uniform priors within the parameter ranges permitted by the models.
Given that we have only two free parameters in our model (specifically, $\Gamma$ and $n_{\rm H}^{\rm local}$), we have conducted the MCMC chain with 8 walkers, and we have run it for a total of $5\times10^5$ steps to ensure chain convergence. We have discarded the initial $10^5$ steps of the chain, commonly referred to as the `burn-in' phase, to eliminate any influence of the initial starting points on our results. All the uncertainties in the parameters are quoted at the 90\% confidence level.

The distribution of $n_{\rm H}^{\rm local}$ is shown in Figure \ref{fig:local_nh_distr}. 
The local obscuring column densities exceeding $10^{22} \rm cm^{-2}$ in X-ray data typically correspond to X-ray type-II AGNs \citep[][and references therein]{viitanen_et_al_2023}. Our analysis has revealed that only a limited number of AGNs within our sample exhibit such high obscuration levels. Interestingly, we have not observed a clear bimodality in the distribution of column densities, which may be attributed to the relatively small number of AGNs displaying high levels of obscuration in our dataset. Therefore, it is reasonable to conclude that the total population primarily consists of type-I AGNs with varying $n_{\rm H}^{\rm local}$ ranging from $4.23^{+1.93}_{-3.74} \times 10^{19} \, \rm cm^{-2}$ to $6.43^{+3.34}_{-4.33} \times 10^{22} \, \rm cm^{-2} $. 

%%%%%% distribution of intrinsic properties %%%%%%
\begin{figure*}
    \centering
    \begin{subfigure}{0.33\textwidth}
        \centering
        \includegraphics[width=0.9\linewidth, height = 2in]{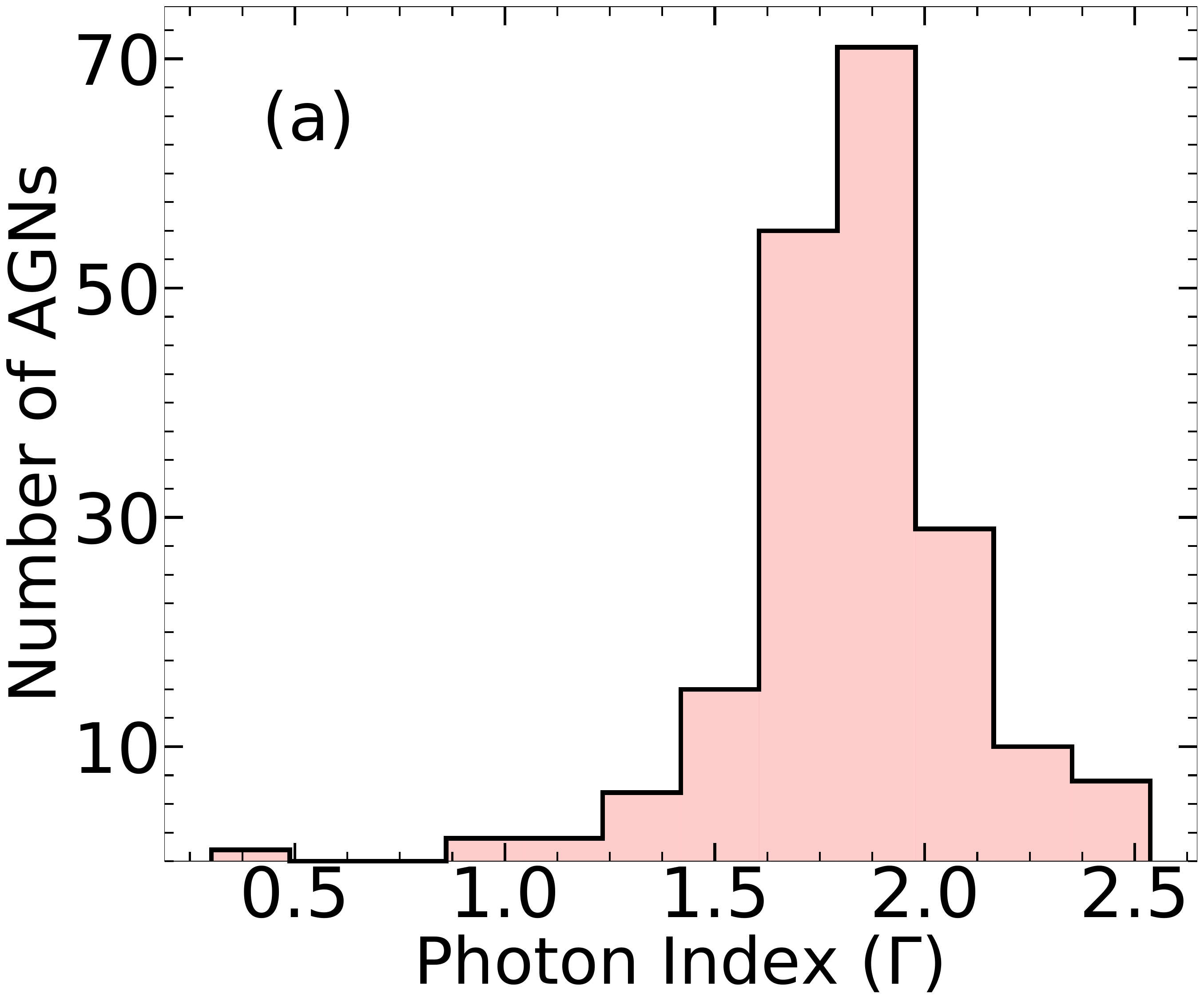}
    \end{subfigure}%
    \centering
    \begin{subfigure}{0.33\textwidth}
        \centering
        \includegraphics[width=0.9\linewidth, height = 2in]{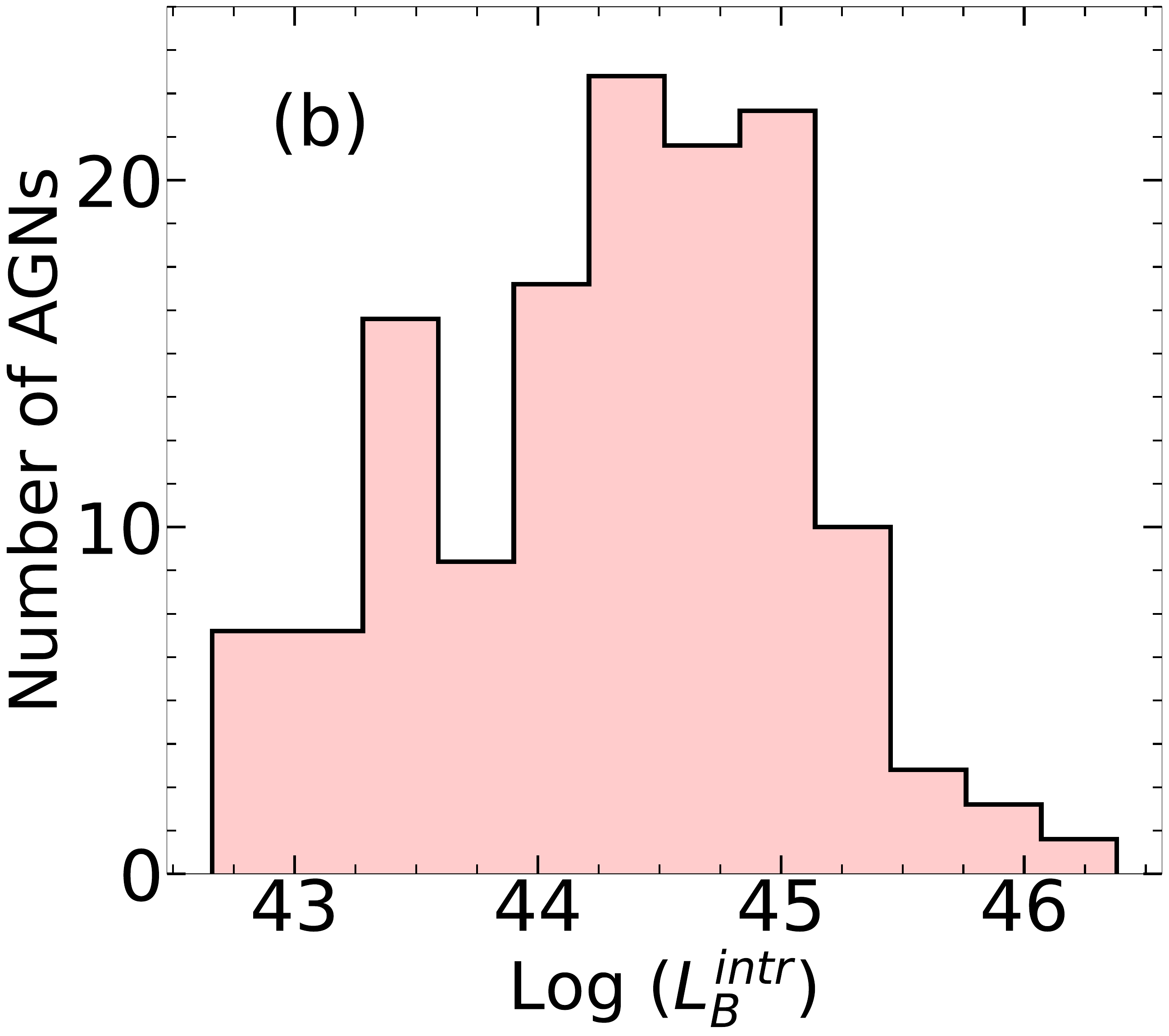}
    \end{subfigure}%
    \centering
    \begin{subfigure}{0.33\textwidth}
        \centering
        \includegraphics[width=0.9\linewidth, height = 2in]{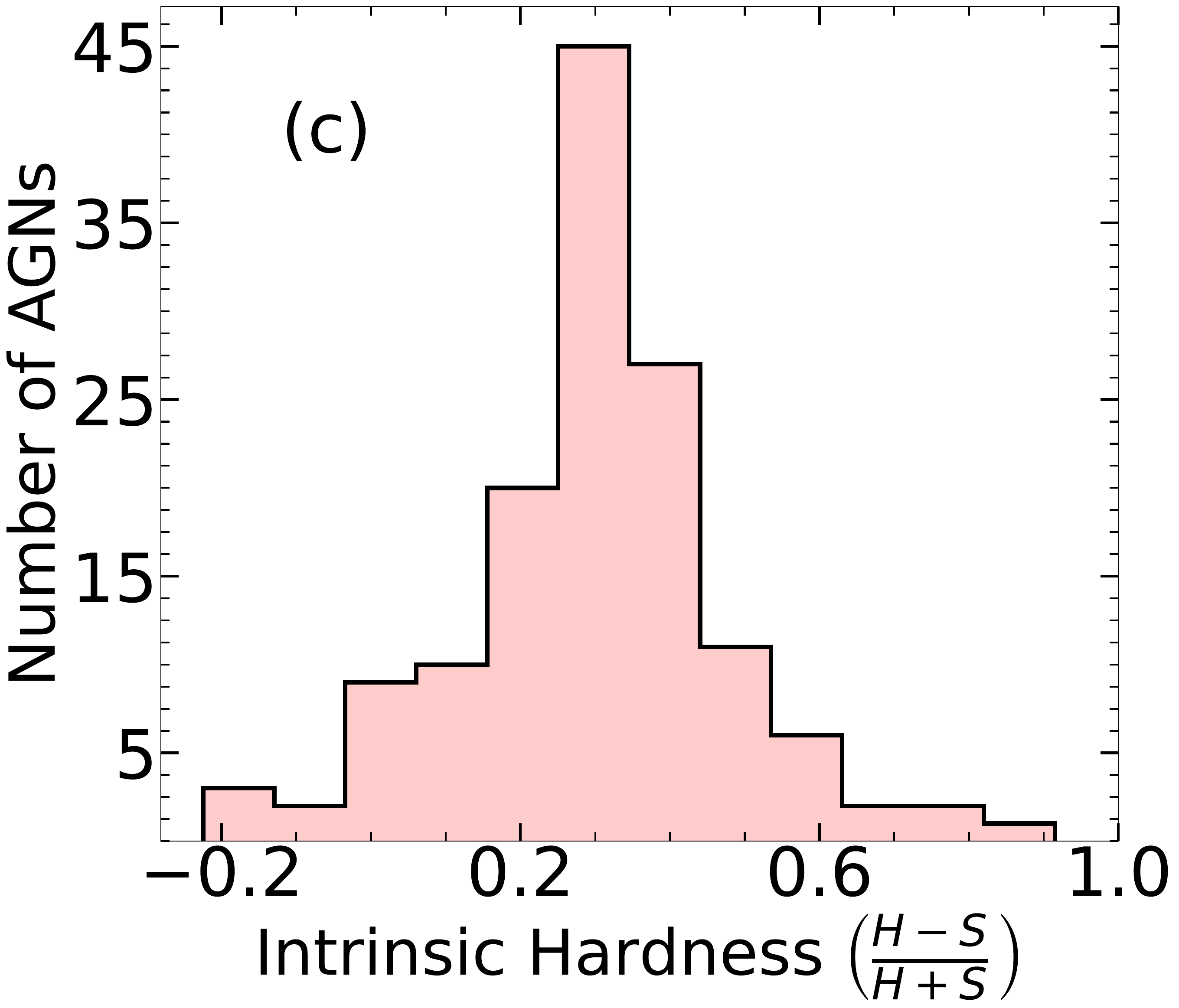}
    \end{subfigure}
    \caption{Distribution of spectral fitting parameters of the new sample AGNs. The distribution of the photon index ($\Gamma$) is shown in (a), whereas the distribution of broadband (B) intrinsic luminosity is presented in (b). Panel (c) represents the distribution of intrinsic hardness between the hard band (H) and soft band (S). (see Section \ref{sect:result} and Table~\ref{tab:lum_band}).}
    \label{fig:gamma_lum_hr_distr}
\end{figure*}

%%%%%%%%%%%%%%%%% Spectrum modeling and analysis of previously known AGNs %%%%%%%%%%%%%%%%%%

\subsection{Spectral modelling of known AGNs}
\label{subsect:modeling_known}
A similar methodology, as detailed in Section \ref{subsect:modeling_new_AGNs}, has been adopted for the spectral modelling of the known sample of AGNs. In this case, only 23 AGNs have exhibited some extent of local obscuration and that too not of very high amount. This is expected given that our known sample exclusively comprises Sy1 galaxies, typically exhibiting minimal intrinsic obscuration. Like the sample of new AGNs, a simple absorbed powerlaw component has proven adequate to fully model the spectra in this known sample except for only two AGNs (2CXO J112956.5+364919 and 2CXO J141449.5+361240). These AGNs display an excess emission at the low-energy end of the spectrum, commonly called soft excess. A possible explanation for this feature is the presence of a multi-colour blackbody emission from the inner part of the accretion disc. For low mass AGNs, this feature can be observed owing to the fact that the temperature of the accretion disc $T^4 \propto (L/L_{\rm Edd})/M$ for a standard geometrically thin and optically thick disc as predicted by \cite{shakura_sunyaev_1973, nobikov_thorne_1973}. This relationship implies that AGNs with relatively lower mass black holes and high accretion rates would have higher accretion disc temperatures, leading to the observed soft excess. However, a multi-color blackbody model alone is often insufficient to fully account for the soft excess observed in AGNs.
\cite{bechtold_et_al_1987}, \cite{Done_et_al_2012} argued that a Compton upscattering of the accretion disc's UV/optical photons by an optically thick and cold corona close to the inner disc could produce the observed soft excess. Alternately, \cite{Nan_ding_et_al_2022} showed that the soft excess could be due to a relativistically smeared reflection of the high-energy X-ray photons by the accretion disc.

In our sample of known AGNs, the soft component is modelled using a disc blackbody (\texttt{diskbb}) component. However, \cite{Gierlinski_&_Done_2004} showed that the estimated temperature associated with the soft excess of AGNs is too high as compared to the disc temperature calculated from the Shakura \& Sunyaev model. Therefore, a multi-color blackbody or other models that provide an adequate fit to the soft excess must be considered as phenomenological model \citep{Piconcelli_et_al_2005}. We have adopted either \texttt{tbabs $\times$ ztbabs $\times$ powerlaw} model or \texttt{tbabs $\times$ ztbabs (diskbb + powerlaw)} for spectral fitting of the known AGNs. The two AGNs in our sample which show soft-excess yield diskbb temperatures of $0.05^{+0.04}_{-0.04}$ and $0.14^{+0.05}_{-0.04}$ keV respectively. While the estimation of the first AGN is well within the theoretical limit (few tens of eV as mentioned in \cite{Gierlinski_&_Done_2004}), the later estimation is very high. However, the introduction of \texttt{diskbb} component has improved the fitting statistics.
We have estimated the model parameters and their uncertainty at the 90\% confidence level described in Section \ref{subsect:modeling_new_AGNs}. Like the new AGNs sample, we have used MCMC for constraining the model parameters in the known sample as well. 

%%%%%%%%%%%%%%%%%%%%%%%%%%%%%%%%%%%%%%%%%%%%%%%%%%%%%%%%%%%%%%%%%%%%%%%%%%%%%%%%%%%%%%%%%%%%%%%%%%%%%%%%%%%%%%%%%%%%
                                    %%% Results %%%
%%%%%%%%%%%%%%%%%%%%%%%%%%%%%%%%%%%%%%%%%%%%%%%%%%%%%%%%%%%%%%%%%%%%%%%%%%%%%%%%%%%%%%%%%%%%%%%%%%%%%%%%%%%%%%%%%%%%%%%%%%%%

\section{Results}
\label{sect:result}
As described in Section~\ref{subsect:modeling_new_AGNs}, we have used an absorbed \texttt{powerlaw} for spectral modelling of the new AGNs. The inferred properties of these AGNs are tabulated in Table~\ref{table_appendix}. The best-fit model yields photon indices ($\Gamma$) varying between $0.3^{+0.16}_{-0.14}$ - $2.54^{+0.14}_{-0.13}$ (Figure \ref{fig:gamma_lum_hr_distr}a) for the newly discovered 198 AGNs, among which 194 ($97.98\%$) have $1<\Gamma<2.5$; with an average photon index for the full sample as $1.83^{+0.06}_{-0.03}$.  
The intrinsic luminosity ($L^{\rm intr}_{\rm X}$) corrected for both MW absorption and local obscuration in various rest frame energy bands are computed for all 145 AGNs for which redshift information is available, utilizing the \texttt{clumin} model component.
The subscript `$\rm X$' denotes the science energy bands according to \textit{Chandra} X-ray Center (CXC) as given in Table \ref{tab:lum_band}. For high redshift sources the soft energy band (S) in the rest frame may be outside the {\it Chandra} response bandpass. Therefore we have extended the {\it Chandra} response to much lower energy ($0.05$ keV) using {\it XSPEC} command to calculate the intrinsic luminosity. We present the broadband intrinsic luminosity distribution in Figure \ref{fig:gamma_lum_hr_distr}b. The sample of new AGNs covers a wide range (four orders) of X-ray broadband luminosity from $4.59^{+0.41}_{-0.41} \times 10^{42}\, {\rm erg~s}^{-1}$ to $2.4^{+0.12}_{-0.12} \times 10^{46}\, {\rm erg~s}^{-1}$.
The hardness ratio is defined based on the luminosities in the high energy band (X2) and the low energy band (X1) as,
\begin{equation}
    \rm Hardness = \frac{\textit{L}_{X2} - \textit{L}_{X1}}{\textit{L}_{X2} + \textit{L}_{X1}}
    \label{eq:hr}
\end{equation}%
Intrinsic hardness ratios are calculated using various combinations of intrinsic luminosities in different bands following Equation \ref{eq:hr}.
%%%%%% symbols of luminosity %%%%%%
\begin{table}
    \centering

    \caption{The energy bands defined by CXC and symbols denote luminosity in various bands.}
    \begin{tabular}[width=\linewidth]{lcc}
         Energy&Energy&Intrinsic\\     Bands&Range&Luminosity\\
         \hline
         Soft Band (S)&0.5-1.2 keV&$L^{\rm intr}_{\rm S}$ \\
         Medium Band (M)&1.2-2.0 keV&$L^{\rm intr}_{\rm M}$ \\
         Hard Band (H)&2.0-7.0 keV&$L^{\rm intr}_{\rm H}$ \\
         Broad Band (B)&0.5-7.0 keV&$L^{\rm intr}_{\rm B}$ \\
         \hline
    \end{tabular}
    \label{tab:lum_band}
\end{table}
The distribution of intrinsic hardness between the hard and the soft bands is shown in Figure \ref{fig:gamma_lum_hr_distr}c. The intrinsic hardness ratios between hard and soft bands range from $-0.22^{+0.05}_{-0.05}$ to $0.91^{+0.03}_{-0.03}$, which indicates the presence of a large variety of spectral shapes in our sample set. Search for the presence of correlations between various parameters is performed as well.

%%%%%%%%%%%%%%%%% Inter relation between different parameters %%%%%%%%%%%%%%%%%%
%%%%%% correlation-anti-correlation between different parameters %%%%%%
\begin{figure}
    \centering
    \includegraphics[width=0.9\columnwidth]{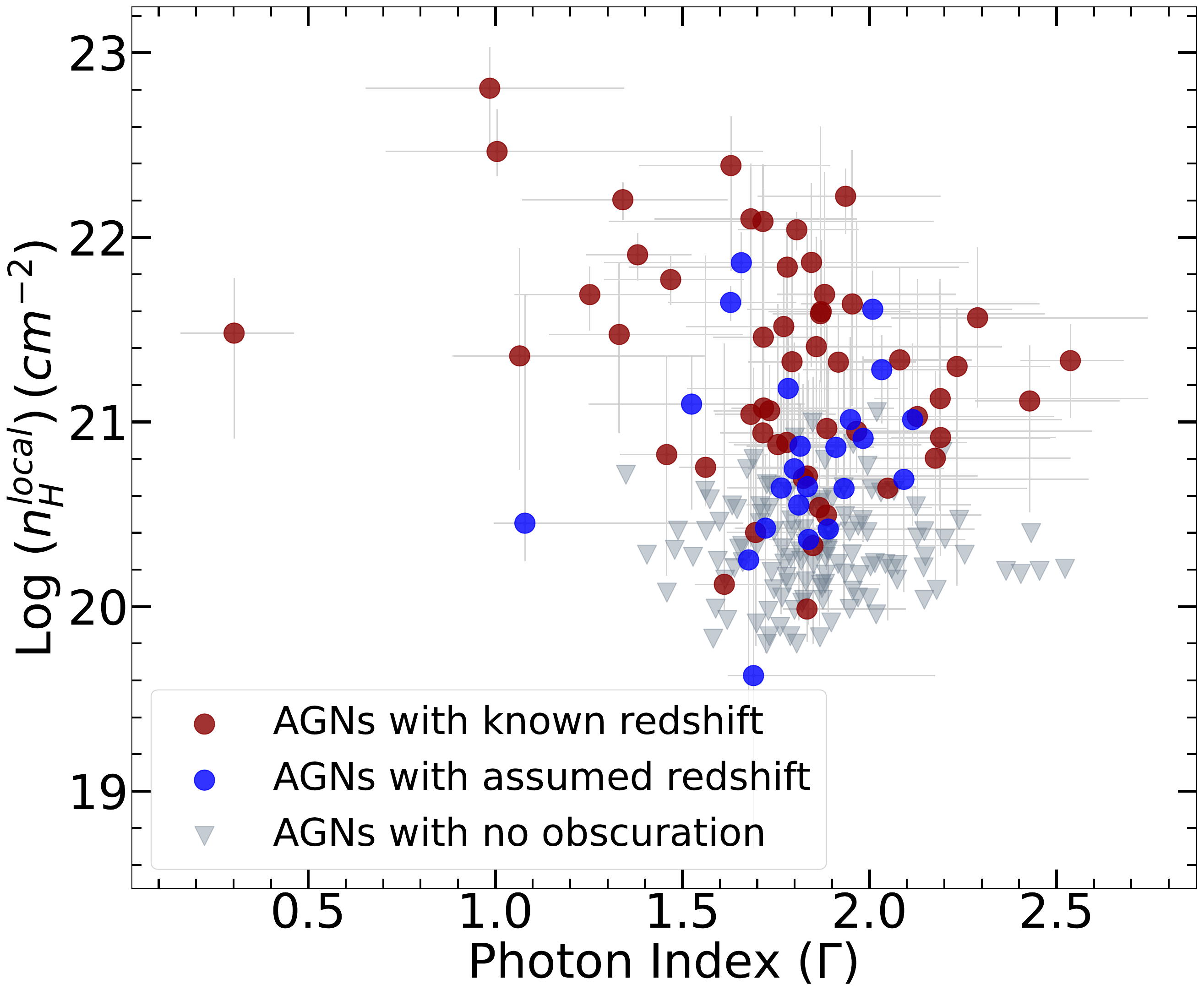}
\caption{Intrinsic  absorption
column density ($n_{\rm H}^{\rm local}$) versus spectral index ($\Gamma$). 
The red data points are AGNs with known redshift, and the blue ones with unknown redshift (assumed $z=0$). The grey triangles represent the upper bound of the column density where no intrinsic obscuration could be found. See Section \ref{subsect:correlation_between_parameters} for details.}
   \label{fig:hardenning_spectra}
\end{figure}

\subsection{Correlation between $\mathbf{\Gamma}$ and $\textbf{\textit{n}}_{\rm \textbf{H}}^{\rm \textbf{local}}$}
\label{subsect:correlation_between_parameters}
An absorbed powerlaw model is used to fit the spectra of new AGNs. Spectra with steeper (shallower) slopes or higher (lower) values of $\Gamma$ correspond to reduced (increased) emissions in the high-energy band, consequently suggesting softer (harder) spectra. Also, intrinsic absorption plays a pivotal role in deciding the observed spectral shape. 
Fig. \ref{fig:hardenning_spectra} shows variations of intrinsic absorption column density ($n_{\rm H}^{\rm local}$) with spectral index ($\Gamma$). Here, we have marked sources with known redshift (red points) and unknown redshift (blue points).
In our sample, 122 AGNs (approximately $62\%$ of the total) exhibit no additional obscuration beyond the Galactic absorption. These sources are marked with grey triangles in Fig. \ref{fig:hardenning_spectra}. The Galactic absorption column densities represent the upper limit for the intrinsic local obscuration of these AGNs. We performed Pearson's correlation test using only the sources with known redshifts (indicated by red data points reflecting the true intrinsic column density), resulting in a correlation coefficient of -0.41, suggesting a weak anti-correlation. However, the degeneracy between these two parameters in the model fitting may also contribute to this anti-correlation.

%%%%%%%%%%%%%%%%% Variation of intrinsic properties with redshift %%%%%%%%%%%%%%%%%%
\subsection{Variation of intrinsic properties with redshift}
\label{subsect:intrinsic_props_vs_z}
\begin{figure*}
    \centering
    \begin{subfigure}{0.5\textwidth}
        \includegraphics[width=0.8\linewidth]{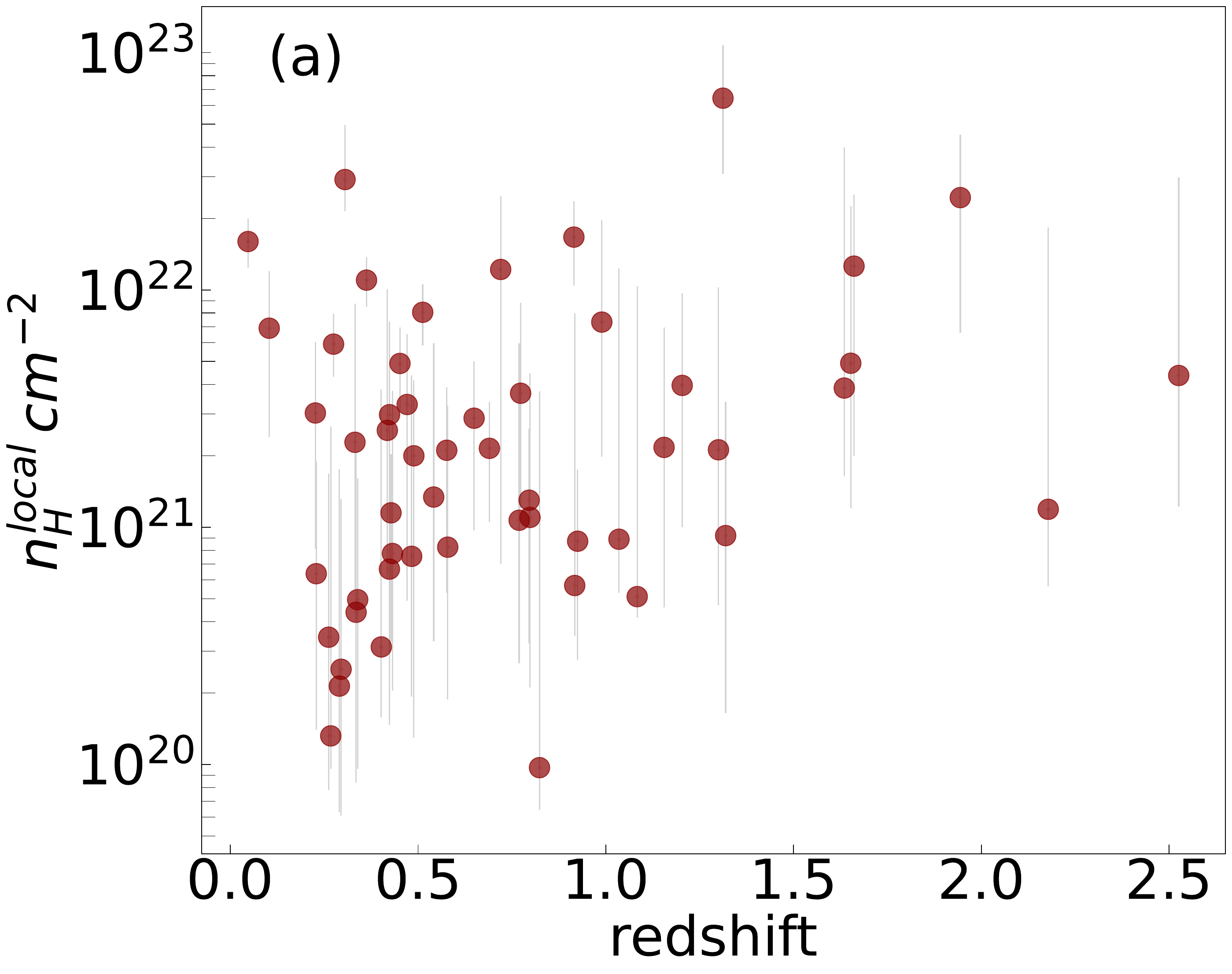}
    \end{subfigure}%
    \centering
    \begin{subfigure}{0.5\textwidth}
        \includegraphics[width=0.8\linewidth]{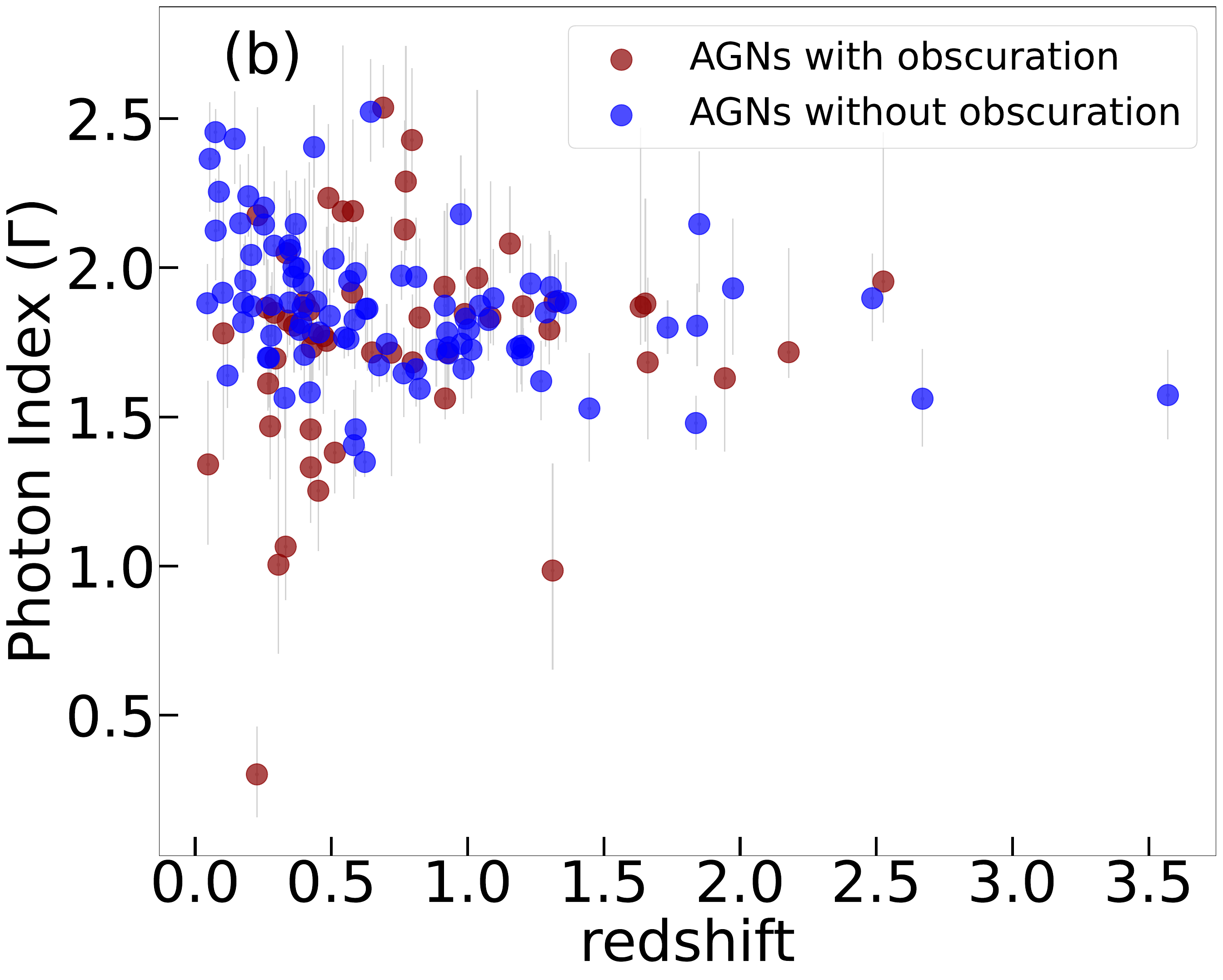}
    \end{subfigure}%
    \caption{Variation of intrinsic parameter with redshift. a) $n_{\rm H}^{\rm local}$ variation with redshift and b) Photon index variation with respect to redshift. The red and blue data points are AGNs with and without some amount of intrinsic absorption respectively. (see Section \ref{subsect:intrinsic_props_vs_z})}
    \label{fig:cosmic_evolution}
\end{figure*}

We have checked how the intrinsic spectral parameters of the AGNs vary with respect to redshift. It can be a very good probe of the cosmic evolution of AGNs where one can use redshift as a proxy for time. Figure \ref{fig:cosmic_evolution}a and \ref{fig:cosmic_evolution}b show the variation of intrinsic absorption column density ($n_{\rm H}^{\rm local}$) and photon index ($\Gamma$) with redshift respectively.
Fig. ~\ref{fig:cosmic_evolution}a illustrates that at low redshifts, Chandra data can detect sources with very low $n_{\rm H}^{\rm local}$. However, this is not feasible at high redshifts because low obscuration primarily affects the softer X-ray emissions (in the AGN's rest frame), which fall outside Chandra's detection range. In contrast, significant obscuration will influence emissions at certain rest-frame energies that remain within Chandra's band after accounting for redshift.

Similarly, the photon index spans over 1.0–2.5 for the nearby sources which becomes narrower in high redshift Universe. This observation may reflect the energy-dependent effective area of \textit{Chandra}. Sources with high intrinsic $\Gamma$ are more luminous in the softer X-ray band, outside \textit{Chandra's} detection range. This could account for the scarcity of softer sources at high redshifts in our sample. Conversely, harder sources should theoretically be detectable; thus, their absence might suggest a lack of harder sources at high redshifts. This trend would become more evident if we had a comparable number of sources in the high-redshift universe.
We indicate AGNs with and without intrinsic obscuration separately using red and blue data points in Fig. ~\ref{fig:cosmic_evolution}b. Some AGNs with obscuration have a lower value of $\Gamma$. However, sources without obscuration do not show very low $\Gamma$.

%%%%%%%%%%%%%%%%% Hardness-Luminosity Diagram %%%%%%%%%%%%%%%%%%

\subsection{Hardness - Luminosity Diagram}
\label{subsect:hld}

\begin{figure}
    \centering
    \includegraphics[width=0.45\textwidth]{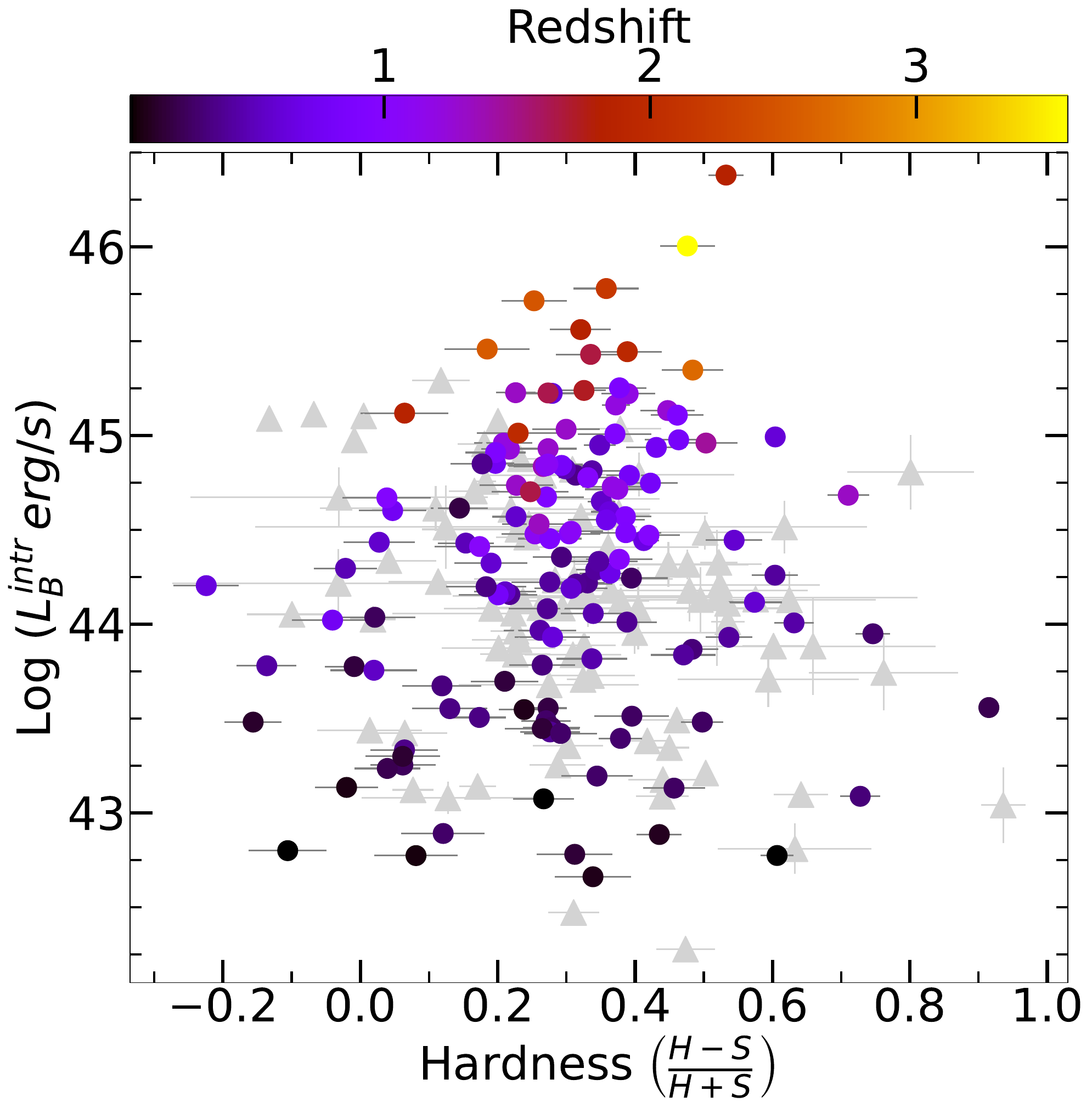}
    \caption{Hardness Luminosity Diagram (HLD) showing the variation of the intrinsic broadband (B) luminosity with intrinsic hardness between hard band (H) and soft band (S). The solid circles denote the newly discovered AGNs, with their respective redshift information depicted using the colour code. The grey triangles represent the known AGNs (see Section \ref{subsect:hld}).}
    \label{fig:HLD}
\end{figure}

Figure \ref{fig:HLD} shows the variation of broadband intrinsic luminosity ($L_{\rm B}^{\rm intr}$) with hardness (HLD) in the H and S bands. A colour-coded scale is provided at the top of the figures to represent the redshift of the new AGNs. Figure \ref{fig:HLD} reveals that the local universe is characterized by a mixture of both harder and softer sources (most of the sources have a hardness > 0). However, in the distant universe ($z > 0.7$), the distribution becomes narrower. This is consistent with the behaviour observed in Section ~\ref{subsect:intrinsic_props_vs_z}, which indicates a relatively higher abundance of soft AGNs compared to hard ones at high redshifts.
Although this effect might be due to the limited number of sources at higher redshifts, it could also suggest an overall cosmic evolution of AGNs, if such evolution exists. To investigate this, we conducted simulations across three small redshift bins ($\Delta z = 0.2$), within which we did not expect significant cosmological evolution, and created 50 sources in each bin. The distribution of sources across these redshift bins shows neither a preferential narrowness at high redshifts nor any clear evolutionary trend over cosmic time (see Figure \ref{fig:simulated_result}). Further details on the simulation can be found in Appendix \ref{appendixA}.

As previously mentioned, we also select a sample set of 91 known AGNs for comparing our results. We have examined the same parameter spaces for these AGNs and the hardness-luminosity space distribution is overplotted using grey triangles with our existing results in Figure~\ref{fig:HLD}.
The overall trends in both sample sets match quite well.

%%%%%%%%%%%%%%%%%%%%%%%%%%%%%%%%%%%%%%%%%%%%%%%%%%%%%%%%%%%%%%%%%%%%%%%%%%%%%%%%%%%%%%%%%%%%%%%%%%%%%%%%%%%%%%%%%%%%%%%%%%%%
                                                    % Discussion%
%%%%%%%%%%%%%%%%%%%%%%%%%%%%%%%%%%%%%%%%%%%%%%%%%%%%%%%%%%%%%%%%%%%%%%%%%%%%%%%%%%%%%%%%%%%%%%%%%%%%%%%%%%%%%%%%%%%%%%%%%%%%

\section{Discussion}
\label{sect:discussion}
Studying the X-ray properties of recently discovered AGNs is important to expand our horizon of knowledge about these objects and their X-ray production mechanisms. In the vast realm of observed AGNs, a smaller subset of newly identified AGNs serves as a valuable probe for uncovering important aspects of their nature. 

The redshift distribution (Figure~\ref{fig:redshift_distr}) of the new AGNs shows a peak $z\sim 0.5$. There are more number of AGNs without or very little obscuration (blue line in Figure~\ref{fig:redshift_distr}) as compared to highly obscured AGNs (red line in Figure~\ref{fig:redshift_distr}) because they are preferably detected due to lower X-ray absorption. Based on the analysis of intrinsic obscuration (Figure~\ref{fig:local_nh_distr}), the distribution peaks at $\sim 10^{21} \rm cm^{-2}$ and most of the sources ($\sim 96\%$) have $n_H^{\rm local} < 10^{22} \rm cm^{-2}$.  Therefore, it is evident from Figure~\ref{fig:redshift_distr}, \ref{fig:local_nh_distr} that our sample of 198 newly identified AGNs predominantly falls into the type-I category. Figure \ref{fig:gamma_lum_hr_distr}a,b demonstrates that, these AGNs span a broad range of X-ray luminosity and possess diverse intrinsic spectral shapes. The photon indices of these AGNs are found within a range consistent with various prior studies \citep[e.g.][]{Reeves_&_Turner_2000, Dewangan_et_al2002, Zadorozhna_et_al2021}.
The majority of the sources exhibit intrinsic hardness in the H and S bands within the range of 0.2 to 0.5 (Figure~\ref{fig:gamma_lum_hr_distr}c), suggesting sources tend to be predominantly characterized by hard X-ray spectra. 

Low redshift AGNs are observed to have a wide range of $n_{\rm H}^{\rm local}$ (Figure~\ref{fig:cosmic_evolution}a ) and the distribution tends toward higher $n_{\rm H}^{\rm local}$. 
Low absorption in high-redshift sources is likely to be missed by \textit{Chandra} due to the effects of redshift transformation (see Section~\ref{subsect:intrinsic_props_vs_z}). 

For most of the nearby sources with redshifts below 0.7, the photon indices fall in the range of 1.0 to 2.5 (Figure~\ref{fig:cosmic_evolution}b), while for sources with $z>1.0$, photon indices are confined to a narrower range of 1.5 to 2.0. This pattern may signify a spectral evolution of these sources over cosmic timescales, reflecting changes in their X-ray emission mechanisms and properties. As discussed in Section~\ref{subsect:intrinsic_props_vs_z}, this may suggest a higher abundance of softer sources at high redshifts. Recently, \cite{Zappacosta_et_al_2023} reported AGNs with steep power-law spectra at very high redshifts ($z > 6$), suggesting potential evolution in AGNs, during the Epoch of Reionization. This could involve changes in coronal properties over time or variations in the coupling between the corona and the accretion disk across different redshifts. \cite{Shehata_et_al_2021_gamma_vs_z} examined the X-ray spectral index of 1,280 sources using a power-law model across a broad redshift range and observed an anti-correlation between $\Gamma$ and $z$ in the \textit{XMM-Newton} data. However, addition of soft excess or reflection components in their spectral modelling accounts for any possible cosmological evolution of spectral index. 

The HLD shows (Figure~\ref{fig:HLD}) interesting results, validated using a sample of known type-I AGNs obtained from VERONCAT \citep{veroncat_13th_ed_2010}. The HLD can also serve as a probe of AGN evolution over cosmic time. For X-ray Binaries, Hardness Intensity Diagrams (HID) are extensively used to study spectral state evolution \citep{xrb1} of transient events. In our case, we can use the HLD to explore the long-term evolution of AGNs, with redshift serving as a proxy for time. In conducting this study, it is essential to assume that all the AGNs within our sample originated at a similar time. In our sample, the redshifts range from 0.044 to 3.57, corresponding to ages ranging from 1.78 Gyr to 13.12 Gyr, as calculated based on \cite{cosmo_cal}. 
We can regard the distant AGNs as the younger counterparts of the AGNs found in the local universe. With this perspective in mind, we understand from Figure~\ref{fig:HLD} that the X-ray emission from AGNs tends to become less luminous as time progresses. However, the highest luminosity does not occur at the highest redshift but at a specific critical redshift, denoted as $z_c=1.84$. 
The sample size of AGNs at high redshift is limited, preventing an extensive study of X-ray Luminosity Functions (XLF), a commonly employed method for investigating AGN evolution. Some studies have indicated a dependence of the XLF on redshift \citep{xlf_with_z1, xlf_with_z2}. \cite{5-10-xlf} showed that the XLF evolves following a luminosity-dependent density evolution model with a critical redshift, $z_c$, above which the evolutionary patterns change. Their analysis of sources observed by CXO estimated a critical redshift of $z_c = 1.99 \pm 0.21$, close to our identified critical redshift.

Our selection of known AGNs with optical classifications as Sy1 galaxies is based on the unification model \citep[e.g.][etc.]{Antonucci_1993, Urry_and_Padovani_1995, Bianchi_et_al_2012}. According to this framework, optical Type I AGNs are expected to align with X-ray Type I AGNs. However, 3-17 \% of AGNs exhibit inconsistencies between their optical and X-ray classifications \citep{Ordovas_et_al_2017}. The spectral modelling (see Section~\ref{subsect:modeling_known}) of all the known AGNs consistently reveals minimal intrinsic obscuration ($\sim 10^{18}$ to $1.8^{0.49}_{-0.08} \times 10^{22}$ with 89 out of 91 AGNs having $n_{\rm H}^{\rm local} < 10^{22} \rm cm^{-2}$), thereby confirming their classification as Type I sources. The known AGN sample comprises sources exhibiting powerlaw emission dominantly ($0.13^{+0.54}_{-0.57} \leq \Gamma \leq 2.49^{+0.27}_{-0.24}$), only except for two sources with soft excess (phenomenologically modelled using \texttt{diskbb}). However, all the newly identified AGNs display the presence of only the powerlaw component.
As discussed in Section~\ref{subsect:modeling_known}, it is possible that AGNs with lower black hole masses and high accretion rates could be linked to the observed soft excess phenomenon. Narrow-line Seyfert 1 galaxy (NLS1) \citep{Osterbrock_&_Pogge_1985, Veron-Cetty_et_al_2001}, could be a suitable candidate group, as they predominantly exhibit soft-excess emissions. We have searched the existing literature to determine whether these 2 sources belong to the NLS1 subclass. 
2CXO J112956.5+364919 has no subclassification reported. The other one, 2CXO J141449.5+361240 can be a NLS1 based on the FWHM of $H_{\beta}$ line in its optical spectrum. The broad component of this line exhibits $1694 \pm 194 \rm km/s$ of FWHM \citep{Rakshit_et_al_2017, Ojha_et_al_2020}. Also, the accretion disc temperature estimated from the spectral fitting of these sources is comparable with the typical disc temperature estimations of NLS1 galaxies \citep[e.g.][]{Leighly_1999} although, in our study, the exact physical condition is not important.

%%%%%%%%%%%%%%%%%%%%%%%%%%%%%%%%%%%%%%%%%%%%%%%%%%%%%%%%%%%%%%%%%%%%%%%%%%%%%%%%%%%%%%%%%%%%%%%%%%%%%%%%%%%%%%%%%%%%%%%%%%%%
                                                    % Summary %
%%%%%%%%%%%%%%%%%%%%%%%%%%%%%%%%%%%%%%%%%%%%%%%%%%%%%%%%%%%%%%%%%%%%%%%%%%%%%%%%%%%%%%%%%%%%%%%%%%%%%%%%%%%%%%%%%%%%%%%%%%%%

\section{Summary}
\label{sect:summary_and_conclusion}
This paper presents the X-ray spectral properties of 198 recently discovered AGNs. All of these new AGNs have been modelled using an absorbed \texttt{powerlaw} model with $\Gamma$ ranging between $0.3^{+0.16}_{-0.14}$ and $2.54^{+0.14}_{-0.13}$. A significant majority, approximately 97.98\%, fall within the range of $1 < \Gamma < 2.5$. 

There are significantly high numbers (122 out of 198.) of sources without any local obscuration in the sample of new AGNs. Even for other sources in the sample, the local obscuration is on the lower side, peaking at $\sim 10^{21} \rm cm^{-2}$ and therefore the new sample of AGNs represents the type-1 category.

The lack of sources with lower obscuration at high redshifts is likely not an intrinsic signature but rather the result of obscuration effects being outside \textit{Chandra}'s bandpass. 
Additionally, a narrower distribution of spectral indices is observed for high-redshift objects compared to those at lower redshifts. This may be attributed to the redshifting of soft excess and reflection bumps. However, an intrinsic scarcity of hard sources at high redshift cannot be ruled out. 
Also, we find that the intrinsic spectral signature of the AGNs did not reveal any significant correlation $n_H^{\rm local}$.

Exploring HLD for these new AGNs, concludes similar behaviour. 
Additionally, we have noted that the maximum intrinsic luminosity within our AGN sample corresponds to a redshift of 1.84. This could potentially be attributed to the luminosity-dependent density evolution of XLF for AGNs, which exhibits a change in behaviour around this specific redshift.

Furthermore, we have conducted a comparative analysis with a sample of known type-I AGNs. While all known AGNs are modelled using only the absorbed \texttt{powerlaw} component, except two of them exhibit an additional soft component modelled by the \texttt{diskbb} component. Nevertheless, these known AGNs reproduce the observed results quite satisfactorily.

%%%%%%%%%%%%%%%%%%%%%%%%%%%%%%%%%%%%%%%%%%%%%%%%%%%%%%%%%%%%%%%%%%%%%%%%%%%%%%%%%%%%%%%%%%%%%%%%%%%%%%%%%%%%%%%%%%%%%%%%%%%%
                                                    % Data Availability %
%%%%%%%%%%%%%%%%%%%%%%%%%%%%%%%%%%%%%%%%%%%%%%%%%%%%%%%%%%%%%%%%%%%%%%%%%%%%%%%%%%%%%%%%%%%%%%%%%%%%%%%%%%%%%%%%%%%%%%%%%%%%
%\section*{Acknowledgements}

\section*{Data Availability}
The data used in this paper are publicly available at the \textit{Chandra} Data Archive (\url{https://cxc.cfa.harvard.edu/cda/}). The results of spectral modelling and parameter estimations will be accessible upon request. We will make any other information relevant to this paper available upon request.



\bibliographystyle{mnras}
\bibliography{ref} % if your bibtex file is called ref.bib
%\begin{thebibliography}{99}
%\end{thebibliography}

%%%%%%%%%%%%%%%%%%%%%%%%%%%%%%%%%%%%%%%%%%%%%%%%%%
%%%%%%%%%%%%%%%%% APPENDICES %%%%%%%%%%%%%%%%%%%%%

\clearpage
\appendix
\section{Test of biases in sample selection}
\label{appendixA}

\begin{figure}
    \centering
    \includegraphics[width=0.8\linewidth]{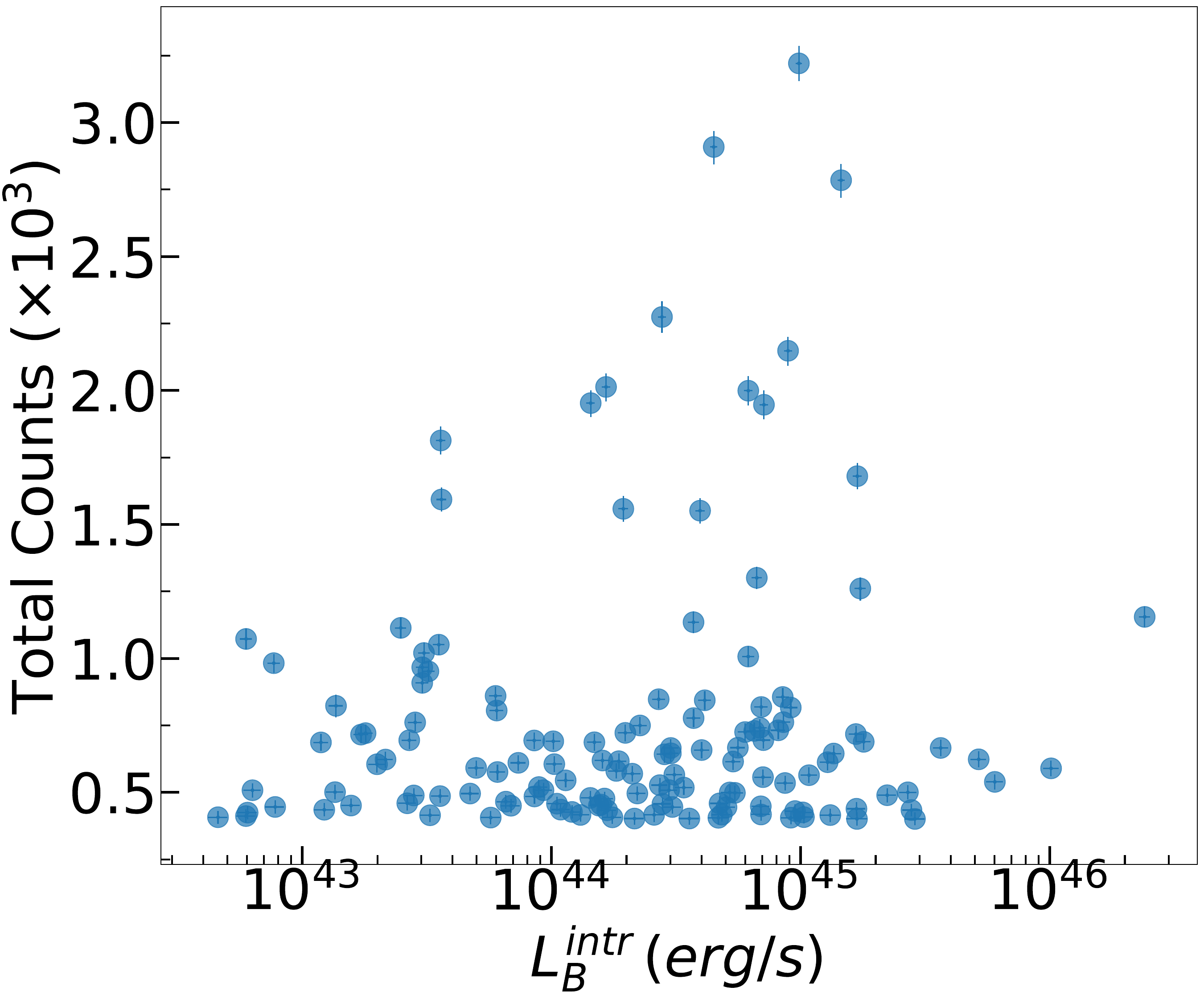}
    \caption{The variation of total counts with broadband (0.5–7.0 keV) intrinsic luminosity.}
    \label{fig:tot_counts_vs_lb_intr}
\end{figure}

It is seen in Figure \ref{fig:redshift_distr} that our sample is mostly populated by AGNs in the local universe, so there is always a possibility of a selection bias while inferring their true nature.
Our source selection is based on total counts rather than flux or count rate, and therefore, the sample should not exhibit any bias toward specific brightness levels. To further examine this, we searched for any correlation between the total counts and luminosity. In Figure \ref{fig:tot_counts_vs_lb_intr}, we plot the total observed counts against the intrinsic luminosity of the sources. Note that the high luminosity sources having systematically lower counts are a signature of bias. However, the total counts within our sample span a broad range and no apparent trend is observed, which rules out the presence of detector sensitivity bias.

\begin{figure}
    \center\centering
    \begin{subfigure}{0.45\textwidth}
        \centering
        \includegraphics[width=0.8\linewidth]{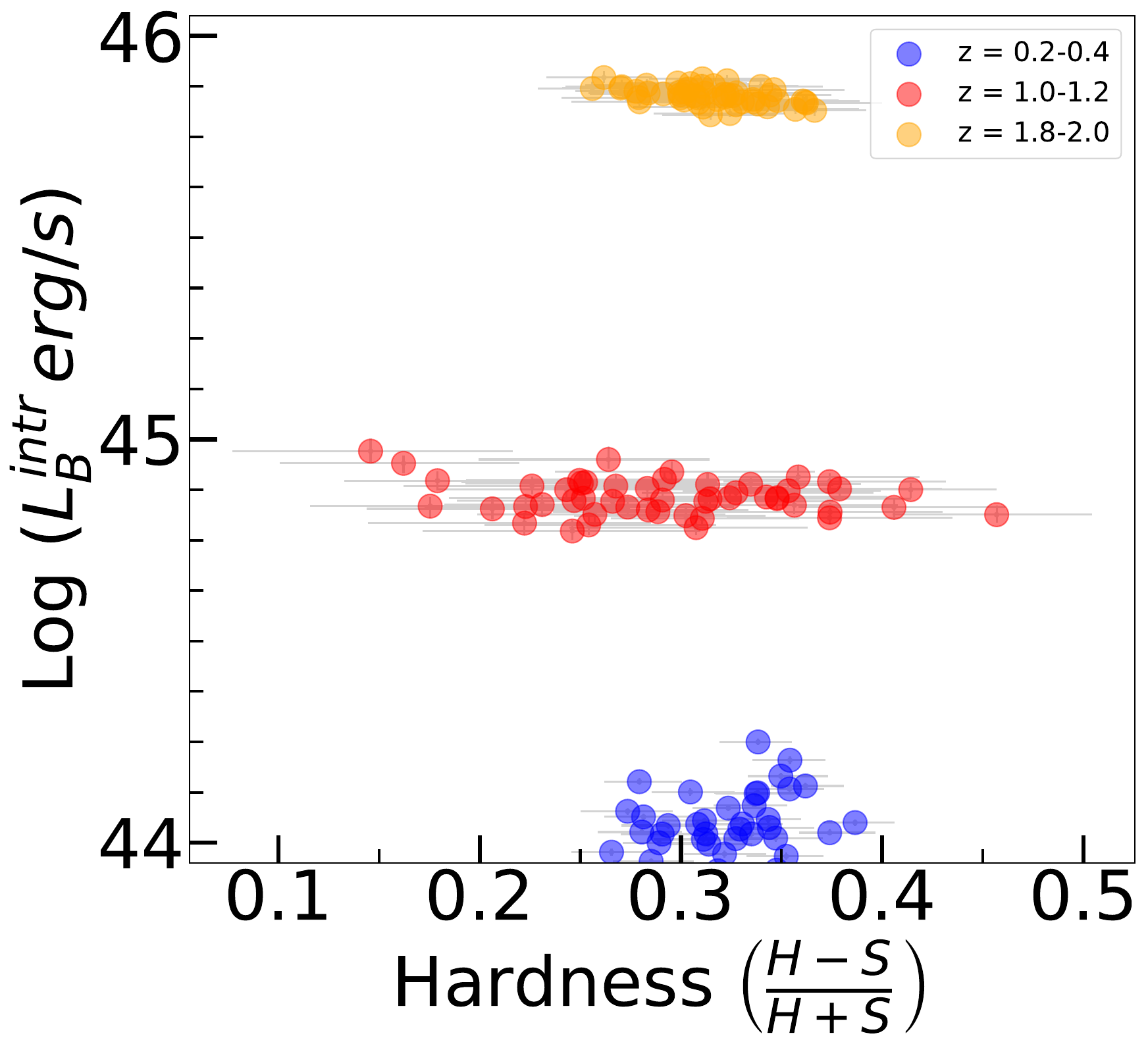}
    % \caption{}
    \end{subfigure}
    \caption{Hardness vs Luminosity diagram of simulated sources for three different redshift bins: z = 0.2–0.4 (blue), z = 1.0–1.2 (red) and z = 1.8–2.0 (yellow).}
    \label{fig:simulated_result}
    \end{figure}

The HLD presented in Figure \ref{fig:HLD} shows a wide span of hardness at lower redshift. In contrast, the distribution becomes narrower at high redshift. This can be an intrinsic property or due to the scarcity of sources at high redshift. To test this, we performed simulations on three different small redshift bins ($\Delta z = 0.2$), where we did not expect any significant cosmological evolution. In each bin, we considered a \textbf{normal} distribution of $\Gamma$ and powerlaw normalization from the observed data and randomly simulated 50 sources. Similarly, we have taken redshifts in each bin to be normally distributed across the bin-range. We estimated the intrinsic hardness in H and S bands, and the results are shown in Figure ~\ref{fig:simulated_result}. The spread of hardness at very low and high redshift is similar and no cosmological evolution is observed in between the redshift bins.

\onecolumn
\section{X-ray properties of the new AGNs}
We list 198 newly identified AGNs from the {\it Chandra} Source Catalog version 2.0 and some of their properties (Table~\ref{table_appendix}).
\renewcommand{\arraystretch}{1.5}
\begin{ThreePartTable}
     \begin{TableNotes}
       \item[a] Obscuring column density local to the AGN.
       \item[b] Powerlaw photon index in $0.5-7.0$ keV.
       \item[c] Intrinsic (both Milky Way and local absorption corrected) broadband ($0.5-7.0$ keV) luminosity.
    \end{TableNotes}  
\centering
\begin{longtable}{|c|c|l|l|l|l|}
    \hline
\endfoot
\caption{Inferred properties of new AGNs, with uncertainties quoted at the $90\%$ confidence level.}\\         
        \hline
        Sl.No.&Source Name&Redshift&$n_{\rm H}^{\rm local}$ ($\rm cm^{-2}$)&$\Gamma$&$L_{\rm B}^{\rm intr}$ ($\rm erg\,s^{-1}$)\\
        \hline
        1&2CXO J212501.2-081328&0.62259&---&$1.35^{+0.05}_{+0.05}$&$9.83^{+0.3}_{-0.3}\times10^{44}$\\
        2&2CXO J161045.0+543612&0.562437&---&$1.76^{+0.06}_{+0.06}$&$4.48^{+0.14}_{-0.14}\times10^{44}$\\
        3&2CXO J105316.7+573550&1.205077&---&$1.73^{+0.07}_{+0.06}$&$1.45^{+0.05}_{-0.05}\times10^{45}$\\
        4&2CXO J095514.5+694735&0.675&---&$1.67^{+0.07}_{+0.07}$&$2.77^{+0.1}_{-0.11}\times10^{44}$\\
        5&2CXO J115937.8+554623&0.546504&---&$1.77^{+0.07}_{+0.07}$&$8.9^{+0.33}_{-0.34}\times10^{44}$\\
        6&2CXO J095416.7+173628&0.383481&---&$1.79^{+0.07}_{+0.07}$&$1.65^{+0.07}_{-0.07}\times10^{44}$\\
        7&2CXO J120921.0-010716&0.36276&$1.1^{+0.28}_{-0.25}\times10^{22}$&$1.81^{+0.17}_{+0.16}$&$6.16^{+0.24}_{-0.24}\times10^{44}$\\
        8&2CXO J002244.4+001825&0.396808&---&$1.95^{+0.08}_{+0.08}$&$1.44^{+0.06}_{-0.06}\times10^{44}$\\
        9&2CXO J122036.3+491150&0.756744&---&$1.97^{+0.08}_{+0.08}$&$7.12^{+0.28}_{-0.28}\times10^{44}$\\
        10&2CXO J101700.7+390432&0.209&---&$1.87^{+0.08}_{+0.08}$&$3.59^{+0.15}_{-0.15}\times10^{43}$\\
        11&2CXO J125639.6+472411&1.319108&$9.22^{+24.5}_{-7.57}\times10^{20}$&$1.89^{+0.16}_{+0.08}$&$1.69^{+0.07}_{-0.07}\times10^{45}$\\
        12&2CXO J124551.0+032128&0.226549&$3.03^{+3.01}_{-2.22}\times10^{21}$&$0.3^{+0.16}_{+0.14}$&$3.62^{+0.16}_{-0.16}\times10^{43}$\\
        13&2CXO J111238.1+132244&0.427953&$1.15^{+0.88}_{-0.82}\times10^{21}$&$1.73^{+0.16}_{+0.15}$&$1.94^{+0.09}_{-0.09}\times10^{44}$\\
        14&2CXO J021820.4-050426&0.64922&$2.88^{+2.13}_{-1.91}\times10^{21}$&$1.72^{+0.14}_{+0.13}$&$3.95^{+0.17}_{-0.17}\times10^{44}$\\
        15&2CXO J003039.5+262056&0.494&---&$1.84^{+0.09}_{+0.09}$&$6.66^{+0.32}_{-0.32}\times10^{44}$\\
        16&2CXO J100434.8+411239&1.7339&---&$1.8^{+0.09}_{+0.09}$&$1.73^{+0.09}_{-0.09}\times10^{45}$\\
        17&2CXO J150424.9+102939&1.837802&---&$1.48^{+0.09}_{+0.09}$&$2.4^{+0.12}_{-0.12}\times10^{46}$\\
        18&2CXO J011305.6+153146&0.576324&$2.11^{+1.77}_{-1.58}\times10^{21}$&$1.92^{+0.17}_{+0.16}$&$3.71^{+0.19}_{-0.19}\times10^{44}$\\
        19&2CXO J095247.0+515053&0.295&$2.52^{+10.6}_{-1.91}\times10^{20}$&$1.7^{+0.21}_{+0.08}$&$2.48^{+0.13}_{-0.13}\times10^{43}$\\
        20&2CXO J111547.4+502405&0.047307&$1.59^{+0.39}_{-0.36}\times10^{22}$&$1.34^{+0.28}_{+0.27}$&$5.94^{+0.32}_{-0.32}\times10^{42}$\\
        21&2CXO J100829.2+072327&0.101082&---&$1.92^{+0.12}_{+0.11}$&$3.53^{+0.19}_{-0.19}\times10^{43}$\\
        22&2CXO J111933.1+212757&0.281371&---&$1.88^{+0.11}_{+0.11}$&$3.08^{+0.17}_{-0.17}\times10^{43}$\\
        23&2CXO J162544.3+154938&0.798005&$1.1^{+3.35}_{-0.89}\times10^{21}$&$1.68^{+0.23}_{+0.1}$&$6.16^{+0.34}_{-0.34}\times10^{44}$\\
        24&2CXO J124828.4+083112&0.118581&---&$1.64^{+0.11}_{+0.11}$&$7.68^{+0.44}_{-0.44}\times10^{42}$\\
        25&2CXO J100434.2+411220&0.275&$5.91^{+2.01}_{-1.6}\times10^{21}$&$1.47^{+0.19}_{+0.18}$&$3.03^{+0.17}_{-0.18}\times10^{43}$\\
        26&2CXO J141531.9+112850&0.360643&---&$2.0^{+0.12}_{+0.12}$&$3.21^{+0.18}_{-0.18}\times10^{43}$\\
        27&2CXO J133140.9-015214&0.145077&---&$2.43^{+0.16}_{+0.15}$&$3.03^{+0.18}_{-0.18}\times10^{43}$\\
        28&2CXO J095003.6+170933&0.194971&---&$2.24^{+0.14}_{+0.14}$&$5.96^{+0.36}_{-0.36}\times10^{43}$\\
        29&2CXO J133240.6+502434&1.231153&---&$1.95^{+0.13}_{+0.13}$&$8.47^{+0.51}_{-0.52}\times10^{44}$\\
        30&2CXO J154107.8+203608&0.508089&---&$2.03^{+0.12}_{+0.11}$&$2.69^{+0.16}_{-0.16}\times10^{44}$\\
        31&2CXO J095048.4+392651&0.205535&---&$2.04^{+0.13}_{+0.12}$&$4.12^{+0.25}_{-0.25}\times10^{44}$\\
        32&2CXO J134452.9+000520&0.086968&---&$2.25^{+0.14}_{+0.13}$&$1.36^{+0.09}_{-0.09}\times10^{43}$\\
        33&2CXO J142738.3+325319&0.823404&$9.7^{+364.0}_{-3.27}\times10^{19}$&$1.83^{+0.26}_{+0.06}$&$6.95^{+0.42}_{-0.43}\times10^{44}$\\
        34&2CXO J103829.9+484925&1.203387&$3.96^{+5.74}_{-2.96}\times10^{21}$&$1.87^{+0.24}_{+0.14}$&$9.12^{+0.56}_{-0.57}\times10^{44}$\\
        35&2CXO J122700.1+215657&0.436337&---&$2.4^{+0.14}_{+0.14}$&$6.02^{+0.38}_{-0.38}\times10^{43}$\\
        36&2CXO J145634.0+222154&0.929287&---&$1.71^{+0.13}_{+0.12}$&$3.72^{+0.24}_{-0.24}\times10^{44}$\\
        37&2CXO J153159.1+242048&0.632057&---&$1.86^{+0.22}_{+0.21}$&$1.67^{+0.14}_{-0.14}\times10^{45}$\\
        38&2CXO J091717.0+415132&1.286578&---&$1.85^{+0.12}_{+0.12}$&$8.52^{+0.55}_{-0.55}\times10^{44}$\\
        39&2CXO J122851.3+033644&0.261967&$3.44^{+13.4}_{-2.66}\times10^{20}$&$1.87^{+0.3}_{+0.1}$&$2.84^{+0.18}_{-0.18}\times10^{43}$\\
        40&2CXO J113201.7+045317&0.339187&$4.95^{+11.1}_{-3.99}\times10^{20}$&$1.82^{+0.26}_{+0.12}$&$2.27^{+0.14}_{-0.15}\times10^{44}$\\
        41&2CXO J115845.4+440000&1.360446&---&$1.88^{+0.14}_{+0.13}$&$6.85^{+0.45}_{-0.45}\times10^{44}$\\
        42&2CXO J141733.7+530404&0.915&$1.67^{+0.7}_{-0.63}\times10^{22}$&$1.94^{+0.25}_{+0.24}$&$8.13^{+0.52}_{-0.53}\times10^{44}$\\
        43&2CXO J104026.8+204544&0.455&---&$1.78^{+0.13}_{+0.13}$&$6.51^{+0.43}_{-0.44}\times10^{44}$\\
        44&2CXO J143832.2+033505&1.004458&---&$1.79^{+0.14}_{+0.14}$&$5.98^{+0.4}_{-0.4}\times10^{44}$\\
        45&2CXO J122226.6+270936&0.488803&$2.0^{+2.16}_{-1.87}\times10^{21}$&$2.23^{+0.25}_{+0.22}$&$1.98^{+0.13}_{-0.13}\times10^{44}$\\
        46&2CXO J113447.3+490133&0.252621&---&$2.14^{+0.13}_{+0.13}$&$1.79^{+0.12}_{-0.12}\times10^{43}$\\
        47&2CXO J235922.6+181130&1.2&---&$1.71^{+0.13}_{+0.12}$&$1.67^{+0.11}_{-0.11}\times10^{45}$\\
        48&2CXO J113421.6+490051&0.228829&$6.37^{+12.5}_{-4.97}\times10^{20}$&$2.18^{+0.36}_{+0.17}$&$1.72^{+0.12}_{-0.12}\times10^{43}$\\
        49&2CXO J123335.0+475800&0.381986&---&$2.0^{+0.14}_{+0.13}$&$7.08^{+0.47}_{-0.47}\times10^{44}$\\
        50&2CXO J105407.2+573524&0.394162&---&$1.87^{+0.15}_{+0.14}$&$2.68^{+0.18}_{-0.18}\times10^{43}$\\
        51&2CXO J125704.2+473815&0.423709&$6.66^{+16.3}_{-5.19}\times10^{20}$&$1.46^{+0.27}_{+0.13}$&$8.52^{+0.58}_{-0.58}\times10^{43}$\\
        52&2CXO J134038.6+402118&0.451831&$4.9^{+2.05}_{-1.77}\times10^{21}$&$1.25^{+0.22}_{+0.2}$&$1.02^{+0.07}_{-0.07}\times10^{44}$\\
        53&2CXO J171518.8+214621&0.885&---&$1.73^{+0.13}_{+0.12}$&$1.79^{+0.12}_{-0.12}\times10^{45}$\\
        54&2CXO J130310.2+333406&0.56541&---&$1.96^{+0.15}_{+0.15}$&$1.49^{+0.1}_{-0.1}\times10^{44}$\\
        55&2CXO J030417.7+002827&0.044431&---&$1.88^{+0.13}_{+0.13}$&$1.19^{+0.08}_{-0.08}\times10^{43}$\\
        56&2CXO J013128.2+003227&0.811122&---&$1.66^{+0.13}_{+0.13}$&$5.59^{+0.38}_{-0.39}\times10^{44}$\\
        57&2CXO J103215.8+574926&1.842331&---&$1.81^{+0.14}_{+0.14}$&$3.64^{+0.26}_{-0.25}\times10^{45}$\\
        58&2CXO J095318.9+515134&1.094636&---&$1.9^{+0.16}_{+0.16}$&$3.01^{+0.22}_{-0.22}\times10^{44}$\\
        59&2CXO J022838.6+003320&0.769203&$1.07^{+4.9}_{-0.8}\times10^{21}$&$2.13^{+0.37}_{+0.12}$&$4.0^{+0.28}_{-0.28}\times10^{44}$\\
        60&2CXO J112014.7+133227&0.992762&---&$1.83^{+0.14}_{+0.14}$&$3.01^{+0.21}_{-0.21}\times10^{44}$\\
        61&2CXO J125456.8+564941&1.269777&---&$1.62^{+0.14}_{+0.13}$&$1.36^{+0.09}_{-0.09}\times10^{45}$\\
        62&2CXO J161126.3+120226&0.925&---&$1.78^{+0.13}_{+0.12}$&$2.84^{+0.2}_{-0.2}\times10^{44}$\\
        63&2CXO J024121.9-080405&2.485&---&$1.9^{+0.15}_{+0.14}$&$5.18^{+0.37}_{-0.37}\times10^{45}$\\
        64&2CXO J143829.1+341513&0.3691&---&$2.15^{+0.14}_{+0.14}$&$2.16^{+0.15}_{-0.15}\times10^{43}$\\
        65&2CXO J113450.1+490327&0.644331&---&$2.52^{+0.18}_{+0.17}$&$1.6^{+0.11}_{-0.11}\times10^{44}$\\
        66&2CXO J105202.7+160247&0.703033&---&$1.74^{+0.13}_{+0.13}$&$1.86^{+0.13}_{-0.14}\times10^{44}$\\
        67&2CXO J122120.1+491845&1.195904&---&$1.74^{+0.14}_{+0.13}$&$5.35^{+0.38}_{-0.38}\times10^{44}$\\
        68&2CXO J085408.8+502312&0.917161&$5.68^{+74.0}_{-2.2}\times10^{20}$&$1.56^{+0.3}_{+0.07}$&$1.28^{+0.09}_{-0.09}\times10^{45}$\\
        69&2CXO J164026.2+533808&0.327972&---&$1.56^{+0.14}_{+0.14}$&$7.36^{+0.52}_{-0.52}\times10^{43}$\\
        70&2CXO J022239.7+014715&0.401&---&$1.71^{+0.14}_{+0.14}$&$1.03^{+0.08}_{-0.07}\times10^{44}$\\
        71&2CXO J155722.1+352928&0.165251&---&$2.15^{+0.2}_{+0.19}$&$1.99^{+0.15}_{-0.15}\times10^{43}$\\
        72&2CXO J022313.3+861913&0.1837&---&$1.96^{+0.15}_{+0.14}$&$4.98^{+0.36}_{-0.36}\times10^{43}$\\
        73&2CXO J221748.2+022010&3.57&---&$1.57^{+0.15}_{+0.15}$&$1.01^{+0.07}_{-0.07}\times10^{46}$\\
        74&2CXO J101039.3-124414&0.424&$2.98^{+4.38}_{-2.11}\times10^{21}$&$1.33^{+0.33}_{+0.19}$&$1.82^{+0.13}_{-0.14}\times10^{44}$\\
        75&2CXO J111709.8+181218&0.34638&---&$1.88^{+0.15}_{+0.14}$&$6.07^{+0.44}_{-0.45}\times10^{43}$\\
        76&2CXO J123725.2+114158&0.5899&---&$1.98^{+0.16}_{+0.15}$&$2.11^{+0.16}_{-0.16}\times10^{44}$\\
        77&2CXO J143823.4+341220&1.0761&---&$1.83^{+0.14}_{+0.14}$&$3.11^{+0.23}_{-0.23}\times10^{44}$\\
        78&2CXO J011835.2-005738&1.3&$2.12^{+8.13}_{-1.65}\times10^{21}$&$1.79^{+0.33}_{+0.12}$&$1.08^{+0.08}_{-0.08}\times10^{45}$\\
        79&2CXO J031846.7-055718&1.045&---&$1.87^{+0.16}_{+0.15}$&$7.06^{+0.54}_{-0.54}\times10^{44}$\\
        80&2CXO J111752.3+174521&0.482966&$7.54^{+36.0}_{-5.61}\times10^{20}$&$1.76^{+0.38}_{+0.12}$&$1.14^{+0.09}_{-0.09}\times10^{44}$\\
        81&2CXO J132118.8+110650&2.178128&$1.19^{+17.1}_{-0.63}\times10^{21}$&$1.72^{+0.35}_{+0.09}$&$6.01^{+0.46}_{-0.46}\times10^{45}$\\
        82&2CXO J085556.1+371342&0.765&---&$1.65^{+0.15}_{+0.15}$&$8.65^{+0.65}_{-0.66}\times10^{44}$\\
        83&2CXO J162720.7+452519&0.578997&$8.24^{+24.3}_{-6.36}\times10^{20}$&$2.19^{+0.31}_{+0.13}$&$2.72^{+0.2}_{-0.2}\times10^{44}$\\
        84&2CXO J115544.5-014740&0.30549&$2.92^{+2.03}_{-0.77}\times10^{22}$&$1.0^{+0.71}_{+0.3}$&$8.9^{+0.71}_{-0.72}\times10^{43}$\\
        85&2CXO J142331.3+240909&1.332896&---&$1.89^{+0.17}_{+0.16}$&$3.39^{+0.28}_{-0.27}\times10^{44}$\\
        86&2CXO J115550.3+232356&0.984723&---&$1.66^{+0.16}_{+0.15}$&$2.96^{+0.23}_{-0.23}\times10^{44}$\\
        87&2CXO J122918.1+034312&0.445&---&$1.89^{+0.17}_{+0.16}$&$9.28^{+0.74}_{-0.74}\times10^{43}$\\
        88&2CXO J142220.1+294255&0.053297&---&$2.37^{+0.19}_{+0.18}$&$6.31^{+0.51}_{-0.51}\times10^{42}$\\
        89&2CXO J161003.1+543627&0.267456&$1.32^{+25.3}_{-0.36}\times10^{20}$&$1.61^{+0.42}_{+0.08}$&$1.35^{+0.11}_{-0.11}\times10^{43}$\\
        90&2CXO J135907.0+470008&1.660953&$1.26^{+1.26}_{-1.06}\times10^{22}$&$1.68^{+0.28}_{+0.26}$&$2.69^{+0.22}_{-0.21}\times10^{45}$\\
        91&2CXO J141648.6+530527&1.181255&---&$1.73^{+0.16}_{+0.15}$&$5.19^{+0.41}_{-0.41}\times10^{44}$\\
        92&2CXO J120154.4+580332&1.305&---&$1.93^{+0.18}_{+0.17}$&$5.44^{+0.43}_{-0.43}\times10^{44}$\\
        93&2CXO J033204.0-273725&1.014&---&$1.73^{+0.17}_{+0.16}$&$2.21^{+0.18}_{-0.18}\times10^{44}$\\
        94&2CXO J111255.5+133206&0.346029&---&$2.08^{+0.18}_{+0.18}$&$4.71^{+0.39}_{-0.39}\times10^{43}$\\
        95&2CXO J150616.1+014131&2.66937&---&$1.56^{+0.17}_{+0.16}$&$2.22^{+0.18}_{-0.18}\times10^{45}$\\
        96&2CXO J121757.8+280513&0.177883&---&$1.88^{+0.2}_{+0.19}$&$2.8^{+0.23}_{-0.23}\times10^{43}$\\
        97&2CXO J120239.7+575249&0.349383&---&$2.06^{+0.17}_{+0.17}$&$3.57^{+0.28}_{-0.28}\times10^{43}$\\
        98&2CXO J114551.5+313349&0.626158&---&$1.86^{+0.19}_{+0.18}$&$8.54^{+0.72}_{-0.72}\times10^{43}$\\
        99&2CXO J120019.0+553348&0.811345&---&$1.97^{+0.2}_{+0.19}$&$1.43^{+0.12}_{-0.12}\times10^{44}$\\
        100&2CXO J141956.6+060626&0.389083&---&$1.82^{+0.17}_{+0.16}$&$1.63^{+0.13}_{-0.14}\times10^{44}$\\
        101&2CXO J104343.1+585535&0.470935&$3.29^{+3.24}_{-2.8}\times10^{21}$&$1.77^{+0.29}_{+0.26}$&$6.56^{+0.53}_{-0.54}\times10^{43}$\\
        102&2CXO J160153.7+431817&0.29042&$2.14^{+15.4}_{-1.51}\times10^{20}$&$1.85^{+0.38}_{+0.1}$&$2.64^{+0.21}_{-0.22}\times10^{43}$\\
        103&2CXO J142814.2+342922&0.9161&---&$1.87^{+0.19}_{+0.18}$&$4.72^{+0.4}_{-0.4}\times10^{44}$\\
        104&2CXO J020446.3-051022&0.773092&$3.67^{+5.17}_{-2.47}\times10^{21}$&$2.29^{+0.45}_{+0.23}$&$1.05^{+0.09}_{-0.09}\times10^{44}$\\
        105&2CXO J124521.8+270755&0.588915&---&$1.46^{+0.16}_{+0.16}$&$2.79^{+0.23}_{-0.23}\times10^{44}$\\
        106&2CXO J160047.6+331311&0.360723&---&$1.97^{+0.18}_{+0.18}$&$1.58^{+0.13}_{-0.14}\times10^{44}$\\
        107&2CXO J115600.1+232156&0.279076&---&$1.77^{+0.17}_{+0.16}$&$1.57^{+0.13}_{-0.13}\times10^{43}$\\
        108&2CXO J093533.7+170243&0.585&---&$1.83^{+0.17}_{+0.17}$&$1.55^{+0.13}_{-0.13}\times10^{44}$\\
        109&2CXO J114837.2+554459&0.420588&---&$1.58^{+0.17}_{+0.17}$&$6.87^{+0.58}_{-0.59}\times10^{43}$\\
        110&2CXO J142845.0+350903&0.9893&$7.32^{+12.4}_{-5.34}\times10^{21}$&$1.84^{+0.42}_{+0.2}$&$6.91^{+0.57}_{-0.57}\times10^{44}$\\
        111&2CXO J140646.6+341528&0.29&---&$2.07^{+0.22}_{+0.2}$&$7.78^{+0.68}_{-0.68}\times10^{42}$\\
        112&2CXO J121515.6+331013&0.925&$8.73^{+8.73}_{-5.98}\times10^{20}$&$1.72^{+0.5}_{+0.11}$&$3.06^{+0.26}_{-0.26}\times10^{44}$\\
        113&2CXO J020456.9-050337&1.65231&$4.91^{+17.7}_{-3.71}\times10^{21}$&$1.88^{+0.35}_{+0.13}$&$5.04^{+0.42}_{-0.43}\times10^{44}$\\
        114&2CXO J142940.7+032126&0.252761&---&$2.2^{+0.2}_{+0.19}$&$1.09^{+0.09}_{-0.09}\times10^{44}$\\
        115&2CXO J020423.9-050619&0.332028&$2.28^{+6.47}_{-1.73}\times10^{21}$&$1.06^{+0.5}_{+0.18}$&$1.22^{+0.11}_{-0.11}\times10^{43}$\\
        116&2CXO J230246.0+084522&1.944&$2.45^{+2.07}_{-1.79}\times10^{22}$&$1.63^{+0.27}_{+0.25}$&$2.78^{+0.24}_{-0.23}\times10^{45}$\\
        117&2CXO J131117.8+215834&0.418149&$2.56^{+7.53}_{-1.96}\times10^{21}$&$1.86^{+0.5}_{+0.17}$&$1.67^{+0.14}_{-0.14}\times10^{44}$\\
        118&2CXO J114325.0+220656&0.824&---&$1.59^{+0.19}_{+0.18}$&$9.49^{+0.83}_{-0.84}\times10^{44}$\\
        119&2CXO J115711.8+082623&0.402057&$3.13^{+34.9}_{-1.55}\times10^{20}$&$1.88^{+0.41}_{+0.1}$&$1.21^{+0.1}_{-0.1}\times10^{44}$\\
        120&2CXO J142359.6+224913&0.930689&---&$1.73^{+0.19}_{+0.18}$&$1.02^{+0.09}_{-0.09}\times10^{45}$\\
        121&2CXO J115542.2+232207&0.176122&---&$1.82^{+0.18}_{+0.17}$&$6.03^{+0.52}_{-0.52}\times10^{42}$\\
        122&2CXO J115504.1+233117&1.083574&$5.1^{+98.5}_{-0.93}\times10^{20}$&$1.83^{+0.46}_{+0.09}$&$6.93^{+0.6}_{-0.6}\times10^{44}$\\
        123&2CXO J144654.6+091801&1.312&$6.43^{+4.33}_{-3.34}\times10^{22}$&$0.99^{+0.36}_{+0.33}$&$4.82^{+0.41}_{-0.42}\times10^{44}$\\
        124&2CXO J121937.8+291012&1.035&$8.9^{+114.0}_{-3.61}\times10^{20}$&$1.97^{+0.63}_{+0.11}$&$2.58^{+0.24}_{-0.25}\times10^{44}$\\
        125&2CXO J100307.8+021135&0.5824&---&$1.41^{+0.19}_{+0.18}$&$1.31^{+0.11}_{-0.11}\times10^{44}$\\
        126&2CXO J002235.9+001850&1.85&---&$2.15^{+0.24}_{+0.23}$&$1.31^{+0.12}_{-0.12}\times10^{45}$\\
        127&2CXO J095331.6+515840&0.270815&---&$1.7^{+0.18}_{+0.17}$&$3.25^{+0.29}_{-0.3}\times10^{43}$\\
        128&2CXO J145453.5+032456&0.075242&---&$2.12^{+0.18}_{+0.17}$&$5.94^{+0.51}_{-0.51}\times10^{42}$\\
        129&2CXO J230257.4+084833&1.974&---&$1.93^{+0.23}_{+0.22}$&$1.03^{+0.09}_{-0.09}\times10^{45}$\\
        130&2CXO J142832.3+004341&0.103634&$6.9^{+5.1}_{-4.5}\times10^{21}$&$1.78^{+0.46}_{+0.42}$&$4.59^{+0.41}_{-0.41}\times10^{42}$\\
        131&2CXO J170036.4+465533&0.266842&---&$1.7^{+0.19}_{+0.18}$&$1.75^{+0.15}_{-0.16}\times10^{44}$\\
        132&2CXO J144654.6+091821&0.541619&$1.34^{+4.61}_{-1.01}\times10^{21}$&$2.19^{+0.56}_{+0.18}$&$5.7^{+0.51}_{-0.51}\times10^{43}$\\
        133&2CXO J122225.4+271223&1.446646&---&$1.53^{+0.19}_{+0.18}$&$9.13^{+0.8}_{-0.8}\times10^{44}$\\
        134&2CXO J130730.3+363248&0.975&---&$2.18^{+0.2}_{+0.19}$&$4.67^{+0.42}_{-0.43}\times10^{44}$\\
        135&2CXO J103512.2+575547&0.72&$1.22^{+1.26}_{-1.15}\times10^{22}$&$1.72^{+0.46}_{+0.41}$&$3.57^{+0.33}_{-0.32}\times10^{44}$\\
        136&2CXO J154341.6+385319&0.431876&$7.75^{+29.8}_{-5.7}\times10^{20}$&$1.78^{+0.48}_{+0.16}$&$2.15^{+0.19}_{-0.2}\times10^{44}$\\
        137&2CXO J091917.7+671531&1.635&$3.86^{+36.0}_{-2.22}\times10^{21}$&$1.87^{+0.6}_{+0.13}$&$1.68^{+0.16}_{-0.16}\times10^{45}$\\
        138&2CXO J153131.5+240333&2.525664&$4.36^{+25.4}_{-3.14}\times10^{21}$&$1.95^{+0.5}_{+0.14}$&$2.87^{+0.26}_{-0.26}\times10^{45}$\\
        139&2CXO J022225.3+422450&---&$2.66^{+4.02}_{-2.1}\times10^{20}$&$1.72^{+0.14}_{+0.08}$&---\\
        140&2CXO J230738.6-224753&---&$7.39^{+5.16}_{-4.82}\times10^{20}$&$1.81^{+0.15}_{+0.14}$&---\\
        141&2CXO J122349.6+072659&---&---&$2.45^{+0.08}_{+0.08}$&---\\
        142&2CXO J085307.3+511747&---&$1.79^{+3.5}_{-1.5}\times10^{20}$&$1.68^{+0.12}_{+0.06}$&---\\
        143&2CXO J120403.3+575748&---&$4.44^{+1.02}_{-0.92}\times10^{21}$&$1.63^{+0.18}_{+0.17}$&---\\
        144&2CXO J091332.2+295546&---&---&$1.79^{+0.09}_{+0.08}$&---\\
        145&2CXO J074421.0+375145&---&$7.29^{+2.68}_{-2.53}\times10^{20}$&$1.91^{+0.1}_{+0.1}$&---\\
        146&2CXO J031215.4+011638&---&---&$2.2^{+0.08}_{+0.08}$&---\\
        147&2CXO J110727.8-052135&---&---&$2.07^{+0.08}_{+0.08}$&---\\
        148&2CXO J091434.7+084351&---&---&$1.9^{+0.11}_{+0.11}$&---\\
        149&2CXO J122700.8+214758&---&---&$1.78^{+0.11}_{+0.1}$&---\\
        150&2CXO J053959.4+771644&---&---&$1.8^{+0.12}_{+0.12}$&---\\
        151&2CXO J161644.6+554638&---&---&$1.75^{+0.07}_{+0.07}$&---\\
        152&2CXO J133044.2+470359&---&$2.15^{+1.23}_{-1.1}\times10^{21}$&$2.54^{+0.14}_{+0.13}$&---\\
        153&2CXO J124525.4+032319&---&$1.92^{+1.04}_{-0.94}\times10^{21}$&$2.03^{+0.23}_{+0.21}$&---\\
        154&2CXO J123059.5+121131&---&$1.3^{+1.3}_{-0.98}\times10^{21}$&$2.43^{+0.24}_{+0.15}$&---\\
        155&2CXO J004929.9+321557&---&---&$1.99^{+0.17}_{+0.16}$&---\\
        156&2CXO J142040.4+414404&---&---&$1.79^{+0.14}_{+0.13}$&---\\
        157&2CXO J125449.1-123059&---&---&$1.93^{+0.14}_{+0.13}$&---\\
        158&2CXO J074423.8+375428&---&---&$1.74^{+0.09}_{+0.09}$&---\\
        159&2CXO J121546.9+601839&---&$4.39^{+9.63}_{-3.48}\times10^{20}$&$1.76^{+0.32}_{+0.14}$&---\\
        160&2CXO J231422.1-424826&---&---&$2.02^{+0.16}_{+0.15}$&---\\
        161&2CXO J234539.0-641242&---&$1.03^{+1.63}_{-0.76}\times10^{21}$&$2.12^{+0.4}_{+0.2}$&---\\
        162&2CXO J171443.2+530927&---&---&$1.95^{+0.16}_{+0.15}$&---\\
        163&2CXO J105245.5+551917&---&$2.63^{+18.8}_{-1.7}\times10^{20}$&$1.89^{+0.39}_{+0.1}$&---\\
        164&2CXO J122658.3+215336&---&$2.31^{+14.5}_{-1.51}\times10^{20}$&$1.84^{+0.42}_{+0.11}$&---\\
        165&2CXO J171456.9+572648&---&---&$1.77^{+0.14}_{+0.14}$&---\\
        166&2CXO J015542.4-554401&---&---&$1.6^{+0.16}_{+0.15}$&---\\
        167&2CXO J225903.6-605929&---&$1.52^{+1.43}_{-1.24}\times10^{21}$&$1.78^{+0.29}_{+0.27}$&---\\
        168&2CXO J160850.8+653220&---&---&$2.13^{+0.16}_{+0.15}$&---\\
        169&2CXO J104807.0+123144&---&$4.38^{+16.3}_{-3.54}\times10^{20}$&$2.05^{+0.28}_{+0.11}$&---\\
        170&2CXO J123702.7+621543&---&$8.05^{+2.52}_{-2.21}\times10^{21}$&$1.38^{+0.14}_{+0.14}$&---\\
        171&2CXO J084032.0+295332&---&$1.03^{+1.85}_{-0.74}\times10^{21}$&$1.95^{+0.41}_{+0.19}$&---\\
        172&2CXO J000313.3+160827&---&$8.15^{+14.5}_{-5.91}\times10^{20}$&$1.98^{+0.5}_{+0.22}$&---\\
        173&2CXO J141528.7-002634&---&$2.17^{+4.77}_{-1.71}\times10^{21}$&$2.08^{+0.19}_{+0.1}$&---\\
        174&2CXO J101720.9+413126&---&---&$1.83^{+0.17}_{+0.16}$&---\\
        175&2CXO J103141.0+350615&---&---&$1.61^{+0.11}_{+0.1}$&---\\
        176&2CXO J111959.7-115610&---&---&$1.69^{+0.16}_{+0.15}$&---\\
        177&2CXO J161152.2+120405&---&---&$2.01^{+0.18}_{+0.18}$&---\\
        178&2CXO J012741.2-061111&---&$4.36^{+17.1}_{-3.38}\times10^{20}$&$1.93^{+0.49}_{+0.15}$&---\\
        179&2CXO J132427.9+044849&---&---&$1.65^{+0.17}_{+0.16}$&---\\
        180&2CXO J044206.7+015851&---&---&$1.85^{+0.18}_{+0.17}$&---\\
        181&2CXO J110557.4-000044&---&$5.58^{+21.4}_{-4.17}\times10^{20}$&$1.8^{+0.4}_{+0.12}$&---\\
        182&2CXO J030905.6+264950&---&---&$2.02^{+0.11}_{+0.11}$&---\\
        183&2CXO J132959.8-020456&---&$4.08^{+2.52}_{-2.23}\times10^{21}$&$2.01^{+0.37}_{+0.34}$&---\\
        184&2CXO J130610.3+035901&---&$7.29^{+3.41}_{-2.98}\times10^{21}$&$1.66^{+0.4}_{+0.37}$&---\\
        185&2CXO J032713.1-132334&---&$4.89^{+19.9}_{-3.69}\times10^{20}$&$2.09^{+0.49}_{+0.14}$&---\\
        186&2CXO J155554.2+663003&---&$4.46^{+12.3}_{-3.55}\times10^{20}$&$1.83^{+0.29}_{+0.12}$&---\\
        187&2CXO J135244.2+312845&---&$4.23^{+193.0}_{-3.74}\times10^{19}$&$1.69^{+0.49}_{+0.07}$&---\\
        188&2CXO J102701.8+174214&---&---&$1.99^{+0.19}_{+0.18}$&---\\
        189&2CXO J132521.2+362912&---&---&$1.82^{+0.17}_{+0.16}$&---\\
        190&2CXO J084030.5+130931&---&$1.25^{+1.03}_{-0.91}\times10^{21}$&$1.52^{+0.3}_{+0.28}$&---\\
        191&2CXO J234525.6-641206&---&---&$1.88^{+0.19}_{+0.18}$&---\\
        192&2CXO J100818.7+114625&---&---&$1.63^{+0.19}_{+0.18}$&---\\
        193&2CXO J114604.4+472641&---&---&$1.49^{+0.17}_{+0.16}$&---\\
        194&2CXO J141612.8+113613&---&---&$2.01^{+0.18}_{+0.17}$&---\\
        195&2CXO J141455.2+361437&---&---&$1.59^{+0.21}_{+0.2}$&---\\
        196&2CXO J032715.3-133147&---&$2.83^{+46.0}_{-1.07}\times10^{20}$&$1.08^{+0.58}_{+0.08}$&---\\
        197&2CXO J163911.9+294248&---&---&$1.89^{+0.23}_{+0.22}$&---\\
        198&2CXO J084023.9+295159&---&$3.55^{+14.9}_{-2.71}\times10^{20}$&$1.81^{+0.46}_{+0.14}$&---\\
    \hline
    \insertTableNotes
    \label{table_appendix}
    \end{longtable}
\end{ThreePartTable}

%%%%%%%%%%%%%%%%%%%%%%%%%%%%%%%%%%%%%%%%%%%%%%%%%%

% Don't change these lines
\bsp	% typesetting comment
\label{lastpage}
\end{document}